\documentclass{jfm}
\usepackage[latin9]{inputenc}
\usepackage{booktabs}
\usepackage{amsmath}
\usepackage{graphicx}

\makeatletter

\providecommand{\tabularnewline}{\\}

\usepackage{epstopdf}
\usepackage{epsfig}

\makeatother

\begin{document}

\title[Dynamics of drop impact]{Dynamics of drop impact on solid surfaces: evolution of impact force
and self-similar spreading}

\author{Leonardo Gordillo\aff{1,2} \corresp{\email{leonardo.gordillo@usach.cl}},
Ting-Pi Sun\aff{1} \and Xiang Cheng\aff{1}\corresp{\email{xcheng@umn.edu}}}

\affiliation{\aff{1}Department of Chemical Engineering and Materials Science,
University of Minnesota, Minneapolis, Minnesota 55455, USA\\
\aff{2}Departamento de F\'isica, Universidad de Santiago de Chile,
Av. Ecuador 3493, Estaci\'on Central, Santiago, Chile}
\maketitle
\begin{abstract}
We investigate the dynamics of drop impacts on dry solid surfaces.
By synchronising high-speed photography with fast force sensing, we
simultaneously measure the temporal evolution of the shape and impact
force of impacting drops over a wide range of Reynolds numbers ($\Rey$).
At high $\Rey$, when inertia dominates the impact processes, we show
that the early-time evolution of impact force follows a square-root
scaling, quantitatively agreeing with a recent self-similar theory.
This observation provides direct experimental evidence on the existence
of upward propagating self-similar pressure fields during the initial impact of
liquid drops at high $\Rey$. When viscous forces gradually set in
with decreasing $\Rey$, we analyse the early-time scaling of the
impact force of viscous drops using a perturbation method. The analysis
quantitatively matches our experiments and successfully predicts the trends of the maximum impact force and the associated peak time with decreasing $\Rey$. Furthermore,
we discuss the influence of viscoelasticity on the temporal signature
of impact forces. Last but not least, we also investigate the spreading
of liquid drops at high $\Rey$ following the initial impact. Particularly,
we find an exact parameter-free self-similar solution for the inertia-driven
drop spreading, which quantitatively predicts the height of spreading
drops at high $\Rey$. The limit of the self-similar approach for drop spreading is also discussed. As such, our study provides a quantitative
understanding of the temporal evolution of impact forces across the
inertial, viscous and viscoelastic regimes and sheds new light on
the self-similar dynamics of drop impact processes. 
\end{abstract}

\section{Introduction}

The elegant and ephemeral dynamics of liquid-drop impacts on solid
surfaces have attracted scientists for generations. Since Worthington's
first sketches \citep{1876RSPS...25..498W,1876RSPS...25..261W}, this
deceivingly simple phenomenon have unfolded into one of the richest
fields in fluid mechanics \citep{Rein:1993m,Yarin:2006al,Visser:2015sm,Josserand:2016jf}.
Thanks to the rapid development of high-speed imaging and numerical
simulation techniques in the last 15 years, a clear picture on liquid-drop
impacts gradually emerges. Different regimes during drop impacts have
been resolved, each describing a specific spatiotemporal feature.
Processes such as lamella ejection and splashing \citep{Xu:2005dl,Riboux:2014gz},
maximum spreading \citep{Roisman:2002iv,Clanet:2004jg,Laan:2014hj},
receding and rebound \citep{Biance:2006hy,Zhao:2015r}, corona fingering
\citep{Krechetnikov:2009gx,Agbaglah:2013jy} and air
cushioning \citep{Driscoll:2011mm,Kolinski:2012fx,Klaseboer:2014ke}
have been extensively studied. Among all these features, the impact
force of liquid drops leads to arguably the most important consequence
of impact events. This mechanical outcome of impacts is directly responsible
for numerous natural and industrial processes including soil erosion
\citep{Nearing:1986da}, the formation of granular craters \citep{Zhao:2015r,Zhao:2015sm}
and atmospheric aerosols \citep{Joung:2015nc} and the damage of engineered
surfaces \citep{Mammitt80,Castano:2010prl}.
The impact force of raindrops is also of vital importance to many
living organisms exposed to the element \citep{Brodie:1951cjb,Dickerson:12pnas,Gart:2015bj}.
Nevertheless, compared with the large number of studies on the morphology
of impacting liquid drops, comparatively fewer experiments have been
conducted to investigate the impact force of liquid drops. Most of
the existing works have focused on the maximum impact force
of liquid drops \citep{Nearing:1986da,Grinspan:2010gt,Li:2014gq,Soto:2014fx,BinZhang:2017ix}.
The temporal evolution of impact forces during impacts remains largely
unexplored.

The complexity of drop-impact dynamics, with the evolution of impact
forces as a specific example, arises from the interplay of various
competing factors and the rapid and continuous change of their relative
importance during a drop impact. Dimensionless numbers such
as Mach (impact velocity/sound speed), Reynolds (inertial/viscous
forces), Weber (inertial/capillary forces) and Froude (inertial/gravity
forces) numbers may change several orders of magnitude in a single drop-impact
event, making it a miniature of many branches of fluid mechanics
\citep{Savic:1955p,Roisman:2009cg,Philippi:2016bg,Wildeman:2016kj}.
In spite of this complexity, pioneering theories have shown
that drop-impact dynamics over a wide range of dimensionless numbers
may be controlled by simple self-similar processes \citep{Roisman:2009cg,Eggers:2010gd,Philippi:2016bg}.
Identifying these self-similar processes will not only reduce mathematical
difficulties at localised spatiotemporal scales, but also bridge separate
impacting regimes into a coherent structure \citep{Barenblatt:1692115}.
Unfortunately, exact or even approximate self-similar solutions are
hard to spot in drop impacts. Most of studies rely on simple dimensional
analyses \citep{Rein:1993m,Yarin:2006al,Josserand:2016jf}, which
are useful in determining asymptotic scaling relations but fail to reveal the
underlying self-similar mechanisms in play.

In this paper, we make a two-fold contribution to understand the self-similar
dynamics of drop impacts. First, we conduct systematic experiments
on the temporal evolution of impact forces over a wide range of Reynolds
numbers ($\Rey$). Built on the recent self-similar theory by Philippi
and co-workers \citep{Philippi:2016bg}, we develop a quantitative
understanding of the early-time scaling of impact forces over five
decades of $\Rey$ across inertial, viscous and viscoelastic regimes.
Through this study, we experimentally verify the existence of an upward
propagating self-similar structure during the initial impact of
liquid drops at high $\Rey$ \citep{Eggers:2010gd,Philippi:2016bg}. Our quantitative analysis on the temporal
variation of impact forces also predicts the maximum impact
force and the associated peak time as a function of $\Rey$, which
have been extensively studied in experiments \citep{Nearing:1986da,Grinspan:2010gt,Li:2014gq,Soto:2014fx,BinZhang:2017ix}. Second,
we generalise the self-similar solution of drop spreading proposed
by Eggers and co-workers \citep{Eggers:2010gd} and find an exact
parameter-free closed-form self-similar solution for inertia-driven
drop spreading following the initial impact. Our exact solution quantitatively
predicts the height of spreading drops at high $\Rey$ and demonstrates
both the advantage and the limit of the self-similar approach in resolving the
dynamics of drop spreading. As such, our experiments on the temporal evolution
of impact forces provide a benchmark for verifying numerical and theoretical
models of drop-impact dynamics. Our theoretical analysis constructs a unifying framework
for understanding the early-time evolution of impact forces in different
impact regimes. In addition, the analytical method for finding
the self-similar solution of drop spreading may also be extended to
other relevant hydrodynamic problems.

\section{Experiments\label{sec:experiments}}

We used a syringe pump to generate quasi-static drops with a fixed
diameter $D=2.2\pm0.1$ mm. The drops were made of silicone oils of
a wide range of viscosities $\nu=10^{-1}-10^{6}\;\mathrm{cSt}$, which
were released from different heights, yielding impact velocities $U_{0}$
ranging from 1.4 up to 3.0 m/s. The drops impacted onto a piezoelectric
force sensor (PCB Piezotronics 106B51), which has a force resolution
$0.3\,\mathrm{mN}$, $50$ times smaller than the inertial force scale
$\rho D^{2}U_{0}^{2}$ ($\rho$ is the density of liquid drops), and
a time resolution on the order of 10 $\mu$s, 100 times faster than
the impact time scale $D/U_{0}$. The sensor has a circular contact
area of diameter 15.0 mm, significantly larger than the maximum spreading diameters of our liquid drops. The force signal passed through a signal
conditioner and was recorded via an oscilloscope. To reduce random
noises and small oscillations in the data, we performed a minimal
data smoothing, where moving averages of three data points, one on
each side of the central value, were taken.

Although the maximum impact force of liquid drops has been measured
in previous studies, the presence of strong resonant ringing and the
abnormal slow decay of impact forces have limited the application
of piezoelectric force sensors in resolving the temporal evolution
of impact forces \citep{Nearing:1986da,Grinspan:2010gt,Li:2014gq,Soto:2014fx,BinZhang:2017ix}.
Here, we solved these problems by targeting the impinging drops slightly
off the centre of the force sensor, which significantly reduced resonant
ringing. Furthermore, to remove the slow decay of force signals at
long times, we chose non-polar liquids, silicone oils, as our liquid
drops, which successfully eliminated dipolar interactions between
impacting drops and the piezoelectric sensor. We directly verified
the accuracy of our experimental method by measuring the impact force
of elastic spheres and by comparing the measured impulse with the momentum of impinging drops. Both measurements quantitatively
agree with theoretical predictons (see appendix~\ref{appN}).

Lastly, we also performed high-speed photography of drop impacts at a
rate of $50\,000\,\mathrm{fps}$ (Photron SA-X2). Triggered by falling
drops through a photo-interrupter, force measurements and high-speed
imaging were synchronised, allowing us to simultaneously probe the
kinematics and dynamics of drop impacts.

\begin{figure}
\begin{centering}
\includegraphics{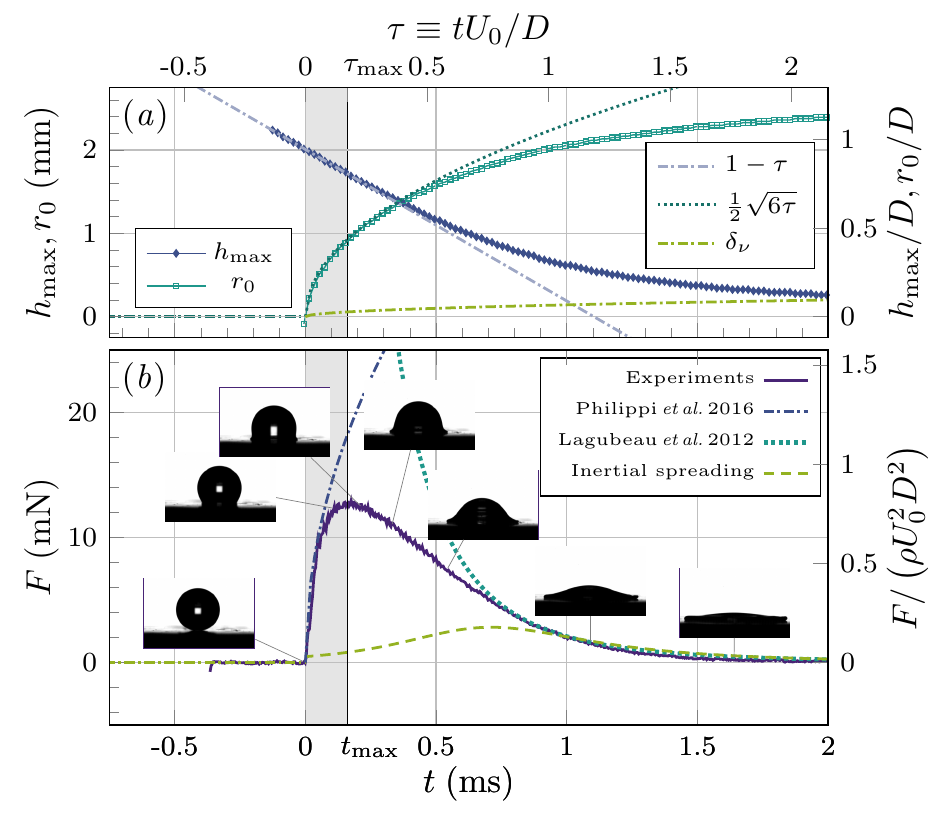} 
\par\end{centering}
\caption{Simultaneous measurement of the kinematics and dynamics of the impact of a liquid drop with
$\nu=20\;\mathrm{cSt}$ and $U_{0}=1.93\,\text{m/s}$ ($\Rey=212$).
(\textit{a}) Temporal evolution of the shape of the impacting
drop, quantified by the height of the drop, $h_{\max}(t)$, and the
radius of the spreading contact line, $r_{0}(t)$. The lower axis
indicates $t$ in unit of milliseconds. The upper axis indicates the
dimensionless time $\tau$. The thickness of the boundary layer, $\delta_{\nu}$,
is calculated and shown by the curved dash-dotted line near the bottom.
The linear dash-dotted line has a slope of $-U_{0}$ (or $-1$ in
the dimensionless form), indicating the trajectory of the drop as
if the impact never occurred. The dotted line indicates the $\sqrt{t}$
scaling of $r_{0}$. (\textit{b}) Temporal
evolution of the impact force of the impacting drop, $F(t)$. The
dash-dotted line on the left is the prediction of the self-similar theory of initial impacts \eqref{eq:Force-1} \citep{Philippi:2016bg}.
The upper dotted line on the right is the prediction of the self-similar
solution of drop spreading by Eggers and coworkers \eqref{eq:impactforceEggers} \citep{Eggers:2010gd,Lagubeau:2012ba}.
The lower dashed line on the right is the prediction of our self-similar
solution of drop spreading. The corresponding snapshots of the impacting drop from high-speed
imaging are shown next to the curve. The regime of the initial impact
is indicated by a shaded area spanning from 0 to $t_{\max}$. A small DC offset
from the force sensor at $t\gg1$ was removed from the raw data.\label{fig1}}
\end{figure}

\section{Results and discussion\label{sec:results}}

Figure~\ref{fig1} shows a representative set of data illustrating
our simultaneous measurement of the shape and impact force of a liquid
drop (see also the Supplementary Video). The impact force displays
a sharp increase upon impact at $\tau=0$, reaches a maximum at $\tau=\tau_{\max}\approx0.2$
and then slowly decays to zero in $\tau\sim2$ (figure~\ref{fig1}\textit{b}),
where $\tau\equiv U_{0}t/D$ is the dimensionless time. Based on the
temporal signature of the impact force, we divide our discussion of
drop-impact dynamics into two parts: (1) the regime
of initial impact before $\tau_{\max}$ and (2) the regime of inertia-driven
spreading at long times after $\tau_{\max}$.

\subsection{Initial impact\label{subsec:initialimpact}}

\subsubsection{Temporal evolution of impact forces at the high-$\Rey$ limit \label{subsec:Experimental-evidence}}

We first investigate the dynamics of liquid drops during the initial
impact near $\tau=0^{+}$ at high Reynolds numbers ($\Rey$), where
$\Rey$ is defined based on the diameter of drops, $\Rey\equiv U_{0}D/\nu$.
In this limit, the impact of liquid drops is dominated by inertia.
Strong pressure gradients develop near the solid surface, which drive
a rapid deformation of the impacting drop and redirect the flow from
the vertical ($z$) to the radial ($r$) direction. In analogy to
the classical impact theory \citep{Wagner:1932h}, simulations and a recent theory have shown that the region of large pressure gradients concentrates within a small volume of the impacting
drop next to the contact area, where self-similar pressure and velocity
fields establish (cf. figure~1 in \citealt{Philippi:2016bg} and
figure~3 in \citealt{Eggers:2010gd}). The relevant length scale of the self-similar fields is given by $\sqrt{U_0Dt}$ (see appendix~\ref{appInertial}). The predicted self-similar pressure gives rise to an instantaneous impact force following
(cf. Eq. (3.39) in \citealt{Philippi:2016bg}) 
\begin{equation}
F\left(t\right)=\frac{3}{2}\sqrt{6}\rho U_{0}^{5/2}D^{3/2}t^{1/2}.\label{eq:Force-1}
\end{equation}
In its dimensionless form, 
\begin{equation}
\widetilde{F}=\frac{3}{2}\sqrt{6}\tau^{1/2},\label{eq:DimensionlessForce-1}
\end{equation}
where $\widetilde{F}\equiv F/\left(\rho D^{2}U_{0}^{2}\right)$ is
the dimensionless force.

A simple scaling argument can be formulated for understanding \eqref{eq:Force-1}.
During the initial impact, the deformation of the drop is limited
within the self-similar high-pressure region. It has been suggested that this high-pressure region occupies a volume with the same radius as the
contact area between the drop and the solid surface \citep{Eggers:2010gd}.
Indeed, previous studies and our experiments have all confirmed that
the radius of the spreading contact line increases as $r_{0}\sim D\sqrt{\tau}\sim\sqrt{U_{0}Dt}$
at short times during initial impact (figure~\ref{fig1}\textit{a}) \citep{Mongruel:2009a,Tabakova:2012s,Riboux:2014gz,Philippi:2016bg},
quantitatively similar to the length scale of the self-similar fields shown above. Thus, we can approximate the volume of the high-pressure region with significant drop deformation as $V\sim\left(U_{0}Dt\right)^{3/2}$. By balancing the impulse of the impact force and the momentum of the deformed drop, the impact force can be simply written as 
\begin{equation}
F\left(t\right)=\frac{\rho VU_{0}}{t}\sim\rho U_{0}^{5/2}D^{3/2}t^{1/2}.\label{eq:scalingForce-1}
\end{equation}

\begin{figure}
\begin{centering}
\includegraphics{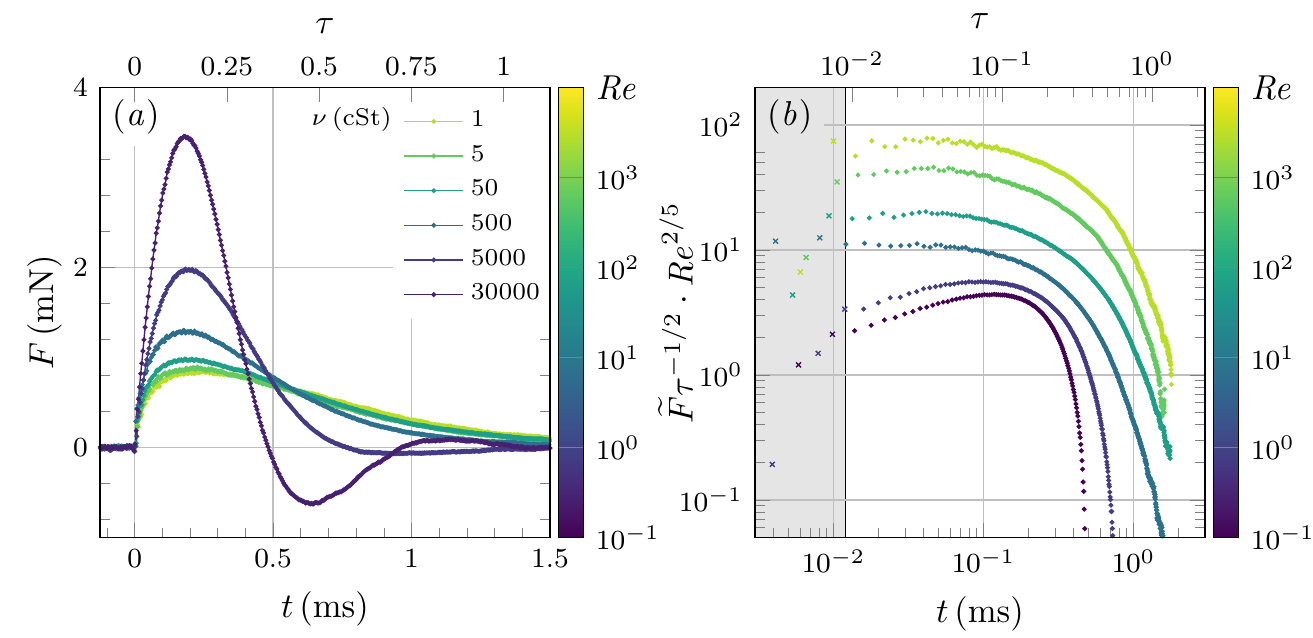} 
\par\end{centering}
\caption{Impact force of liquid drops. (\textit{a}) Temporal evolution of the
impact force of liquid drops at different $\Rey$. In the order of
the maximum impact force, from high to low, the Reynolds numbers of
the curves are $0.10$, $0.72$, $6.90$, $66.18$, $665.52$ and
$3219.29$, respectively. The viscosities of the drops are indicated
in the legend. (\textit{b}) Rescaled dimensionless force, $\widetilde{F}/\tau^{1/2}$,
as a function of time, where $\widetilde{F}$ and $\tau$ are the
dimensionless force and time, respectively. A time-independent factor,
$\mathrm{\Rey}^{2/5}$, is introduced to separate the curves vertically
for clarity. The grey region indicates the rise time of the force
sensor, which sets the time resolution
of our measurements. The Reynolds numbers and viscosities
of the curves are the same as those in (\textit{a}). \label{fig2}}
\end{figure}

We experimentally verify the prediction of the initial-impact self-similar theory
by first plotting the impact forces at different $\Rey$ in a log-log
plot (figure~\ref{fig2}\textit{b}). To reveal the predicted
$t^{1/2}$ scaling at short times, we divide the dimensionless force, $\widetilde{F}$, by
$\tau^{1/2}$. For the sake of clarity, we also multiply the rescaled forces by a time-independent factor, $\Rey$$^{2/5}$, which shifts the curves vertically to avoid overlap.
Figure~\ref{fig2}(\textit{b}) shows that the early-time
evolution of impact forces follows the predicted $\tau^{1/2}$ scaling
at high $\Rey$, where $\widetilde{F}/\tau^{1/2}$ is independent
of $\tau$ for about one decade of time.

\begin{figure}
\begin{centering}
\includegraphics{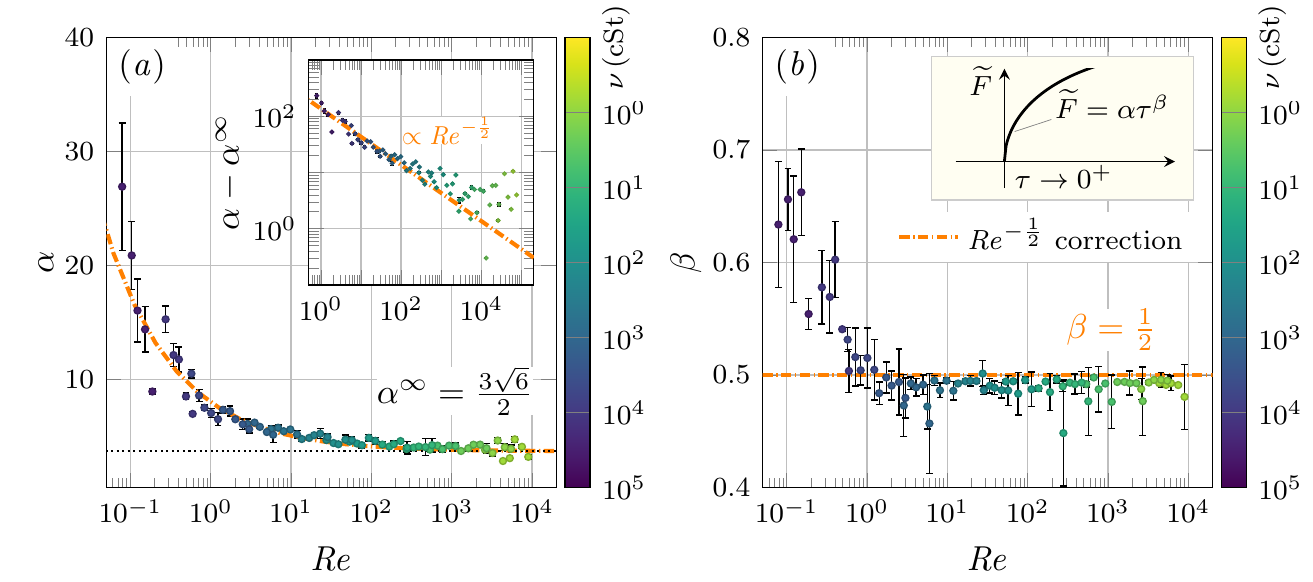} 
\par\end{centering}
\caption{The early-time scaling of impact forces, $\widetilde{F}(\tau)=\alpha\tau^{\beta}$,
near $\tau=0^{+}$ over a wide range of $\Rey$. (\textit{a})\emph{
}The coefficient of the scaling, $\alpha$, and (\textit{b}) the exponent
of the scaling, $\beta$. The range of the power-law fittings spans over one decade
of time starting from $\tau=0^{+}$. The horizontal dotted line in (\textit{a}) indicates
the asymptotic value $\alpha^{\infty}=3\sqrt{6}/2$ predicted by the self-similar theory at the high-$\Rey$ limit \citep{Philippi:2016bg}.
The dash-dotted lines show our model prediction $\alpha(\Rey)$ and
$\beta({\Rey})$ (Eq.~\eqref{eq:viscous_alpha}). The colour bars
on the right of each plot indicate the viscosity of the liquid drops
used in each experiment. The inset of (\textit{a}) shows $\left(\alpha-\alpha^{\infty}\right)$
as a function of $\Rey$ and our theoretical prediction in a log-log
scale. \label{fig3}}
\end{figure}

The data shown in figure~\ref{fig2} represent only a small subset
of our more than 200 independent experimental runs at different $\Rey$.
To quantify all our measurements, we fit $\widetilde{F}$ as a function
of $\tau$ at short times using a power-law dependence, $\widetilde{F}=\alpha\tau^{\beta}$.
The exponent $\beta$ as a function of $\Rey$ for all our measurements
is shown in figure~\ref{fig3}(\textit{b}). $\beta$
reaches a plateau close to $1/2$ when $\Rey>0.7$. The coefficient
$\alpha$ also approaches a constant $\alpha^{\infty}=4.7\pm0.7$,
close to the theoretical prediction $3\sqrt{6}/2$ in \eqref{eq:DimensionlessForce-1},
but only when $\Rey>200$ (figure~\ref{fig3}\textit{a}). Thus, in
combination, our measurements on the early-time evolution of impact
forces quantitatively verify the initial-impact self-similar theory
at high $\Rey$ above $200$.

The existence of upward expanding self-similar fields during the initial
impact of a high-$\Rey$ liquid drop can also be seen from the shape
of the impacting drop. Before the upper bound of the self-similar high-pressure region, marked by the isobar of some preset high pressure, reaches the top surface
of the liquid drop, the motion of the drop apex should remain unchanged
as if the drop had not experienced any impact at all. Such a counterintuitive
hypothesis has indeed already been implied by Worthington's original
sketch \citep{1876RSPS...25..498W,1876RSPS...25..261W} and quantitatively
verified by much more recent simulations \citep{Eggers:2010gd,Roisman:2009ee,Philippi:2016bg}
and experiments \citep{Rioboo:2002mt,Lagubeau:2012ba}. Here, our
simultaneous measurements of the shape and impact force of liquid
drops provide further evidence that this unusual phenomenon arises
from the finite propagation speed of the self-similar fields. As shown
in figure~\ref{fig1}, in the regime where $F(t)$ follows the prediction \eqref{eq:Force-1}, the apex of the drop, $h_{\max}$, keeps traveling
at the initial impact velocity $U_{0}$ without any perceptible
changes. Since the shape of the self-similar region---specifically the isobar of the self-similar pressure field---does not necessarily conform
to the shape of the drop, the pressure field may touch the upper surface
of the drop before reaching the apex. As a result, the impact force
may start to deviate from the prediction of the initial-impact self-similar theory, when $h_{\max}$
is still outside the self-similar region and maintains its constant-velocity
descent. This is indeed consistent with our observations (figure~\ref{fig1}).
The impact force reaches its maximum and begins to decrease before
the apex of the drop shows any clear deviation from $U_{0}$. Thus,
it is more appropriate to use the peak time, $\tau_{\max}$, i.e. the
time when $\widetilde{F}$ reaches the maximum, to mark
the end of the initial impact regime. In practical terms, the maximum
force is easier to identify than the deviation of the drop apex from
its linear descent, which relies on the derivative of $h_{\max}(t)$.

\begin{figure}
\begin{centering}
\includegraphics{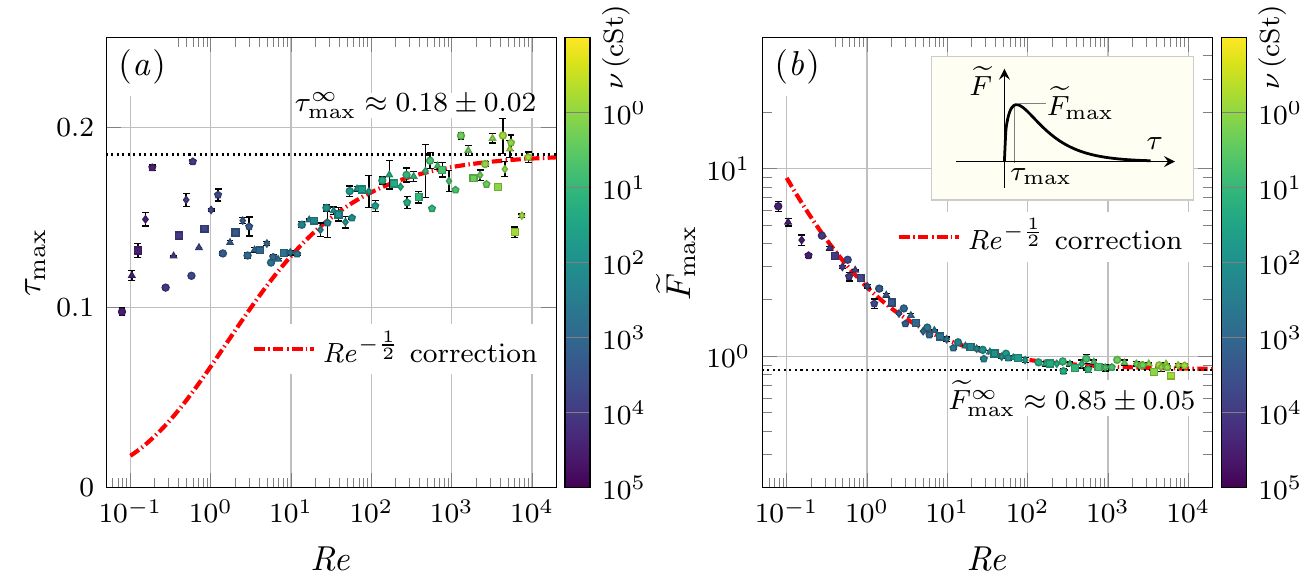} 
\par\end{centering}
\caption{The maximum impact force and the associated peak time. (\textit{a}) shows
the dimensionless peak time, $\tau_{\max}\equiv t_{\max}U_{0}/D$,
and (\textit{b}) shows the dimensionless maximum impact force, $\widetilde{F}_{\max}\equiv F_{\max}/\left(\rho D^{2}U_{0}^{2}\right)$,
over five decades of $\Rey$. The asymptotic values at the high $\Rey$
limit, $\tau_{\max}^{\infty}$ and $\widetilde{F}_{\max}^{\infty}$,
are indicated by the horizontal dashed lines in each plot, which are
obtained experimentally by averaging all the data with $\Rey>10^{3}$.
The dash-dotted lines are our model predictions given by \eqref{eq:tmax1}
and \eqref{eq:fmax}. The colour bars on the right indicate
the viscosity of liquid drops. \label{fig4}}
\end{figure}

The peak time, $\tau_{\max}$, therefore, provides a proper time scale to estimate the average expanding speed of the self-similar fields. A more quantitative analysis of $\tau_{\max}$ based on the propagation of isobars will be provided in $\S$\ref{subsec:FiniteRe} below.
We plot $\tau_{\max}$ and the maximum impact force, $\widetilde{F}_{\max}$,
as a function of $\Rey$ in figures~\ref{fig4}(\textit{a},\textit{b}),
respectively. The value of $\tau_{\max}$ approaches a constant $\tau_{\max}^{\infty}=0.18\pm0.05$
at the high $\Rey$ limit. Since the drop does not deform significantly
during the initial impact, the average expanding speed of the self-similar fields at the high-$\Rey$ limit can be simply estimated as $U_{\mathrm{self-similar}}=D/t_{\max}=U_{0}/\tau_{\max}\approx5.5U_{0}$,
which ranges from 7.7 up to 16.5 m/s in our experiments. Compared with the speed of sound, this relatively small speed demonstrates that the boundaries of the self-similar fields are not shock fronts induced by
the compressibility of liquid drops. Accordingly, $F_{\max}$ should
scale with the inertial force, $\rho D^{2}U_{0}^{2}$, instead of the water-hammer force, $\rho D^{2}cU_{0}$, where $c$ is the speed of sound in the liquid. This argument is indeed supported by both previous
studies \citep{Grinspan:2010gt,Soto:2014fx,Li:2014gq,BinZhang:2017ix} and our experiments (figure~\ref{fig4}\textit{b}).
Thus, the maximum impact force of subsonic liquid drops at high $\Rey$,
relevant to most natural and industrial processes, arises from the
development of upward expanding self-similar pressure fields, rather than water-hammer pressures assumed in several recent studies \citep{Deng:2009t,Kwon:2011hm,Thanh-Vinh:2016n}.

Lastly, it is worth discussing the effect of ambient air on impact
forces. Air cushioning has been the focus of many recent studies (see
\citealt{Josserand:2016jf} and references therein). Although the
ambient air can profoundly affect the dynamics of drop impacts such
as the formation of liquid sheets and splashing \citep{Xu:2005dl,Riboux:2014gz},
the numerical work of Philippi and co-workers showed that the impact
pressure varies smoothly across the air-liquid interface of the air-cushion
layer underneath impacting drops, indicating the transparency
of air cushioning to the impact pressure \citep{Philippi:2016bg}.
Indeed, their study showed that the early-time $t^{1/2}$ scaling
of the impact force is invariant in the presence of ambient air after
they introduced a small time shift, $t^*$, to account for
the delay of the impact moment due to cushioning. We estimate the magnitude of $t^*$ in our experiments as follows. 
By balancing the air pressure with the inertial pressure of the impinging drop, Mani and co-workers showed that the characteristic thickness of the air-cushion layer is $H = R St^{2/3}$, where $R=D/2$ is the radius of the drop and $St = \mu_g/(\rho U_0 R)$ is the inverse of the Stokes number with $\mu_g$ as the air viscosity \citep{Mani:2010m}. Thus, the air-cushion time $t^*$ can be estimated as $t^* = H/U_0$. Using the relevant parameters of our experiments, we find $t^*= 0.12\sim0.41$ $\mu$s, consistent with numerical simulations (cf. figure 2 in \citet{Mani:2010m}). Since $t^*$ is about two orders of magnitude smaller than the temporal resolution of our force sensor (figure~\ref{fig2}\textit{b}), the presence of $t^*$ should not affect the early-time scaling of our experiments. Furthermore, it has been shown that
the impact pressure underneath an impacting drop concentrates near
the moving contact line \citep{Philippi:2016bg}, where air cushioning
is weak or absent \citep{Driscoll:2011mm,Kolinski:2012fx}. Since
the impact force is an integral of the impact pressure over the entire
contact area, which is dominated by the high pressure near the contact
line, air cushioning should not strongly affect the impact force measured
in our experiments. Our measurements indeed show the $t^{1/2}$ scaling
predicted by the initial-impact self-similar theory without ambient air, directly confirming
the weak effect of air cushioning on the early-time scaling of impact
forces. It should be noted that although we cannot directly detect the effect of air cushioning due to the finite time resolution of our force measurements, the existence of the trapped air layer prevents the formation of water-hammer pressures at the very early time of impacts within $t^*$ \citep{Mani:2010m}.   

\subsubsection{Temporal evolution of impact forces at finite $\Rey$\label{subsec:FiniteRe}}

Next, we investigate the early-time scaling of the impact force of
viscous drops, $\widetilde{F}=\alpha\tau^{\beta}$, near $\tau=0^{+}$
at finite $\Rey$. As shown in figure~\ref{fig3}, the coefficient
$\alpha$ starts to deviate from the high-$\Rey$ plateau when $\Rey<200$,
where $\alpha$ increases with decreasing $\Rey$. In contrast, the
exponent $\beta$ maintains at 1/2 until $\Rey\approx0.7$ and then quickly
increases at even lower $\Rey$. In this section, we shall focus on
impact forces, $\widetilde{F}(\tau)$, in the intermediate $\Rey$
regime with $0.7<\Rey<200$ and leave the discussion of $\widetilde{F}(\tau)$
at even lower $\Rey<0.7$ in the next section.

Before delving into rigorous calculations, it is instructive to
consider a simple scaling for impact forces at finite $\Rey$.
At finite $\Rey$, viscous forces cannot be ignored when determining
the dynamics of drop impacts. The distance traveled by the centre
of an impacting drop can be approximated as $d\approx U_{0}t$ at
short times. Based on a simple geometric arguments, the radius of
the contact area between the drop and the solid surface is given by
$r_{0}=\sqrt{dD}=\sqrt{U_{0}Dt}$, as we have already confirmed previously
(figure~\ref{fig1}\textit{a}). Assume the vertical velocity decreases from the impact velocity $U_{0}$ to zero over a length $l$ within the drop above the solid surface. Again, by simply balancing the impulse of the impact force with the change of the momentum of the deformed drop, we have
\begin{equation}
F(t)=\frac{\rho VU_0}{t}\sim\frac{\rho r_0^2 l U_0}{t},\label{eq:ViscousScaling1}
\end{equation}
where $V\sim r_{0}^{2}l$ is the volume of the part of the liquid drop that significantly deforms. At high $\Rey$, $l$ is determined by the self-similar velocity field with $l\sim\sqrt{U_0Dt}$.
Equation~\eqref{eq:ViscousScaling1} restores to the previous scaling \eqref{eq:scalingForce-1}. At finite $\Rey$, the boundary layer developed at the bottom of the impacting drop introduces a new length scale $\delta_{\nu}\approx\sqrt{\nu t}$, which competes with the growth
of the self-similar field that scales as $\sqrt{U_{0}Dt}$. If we set
$l\approx\delta_{\nu}$ in \eqref{eq:ViscousScaling1}, we have $F\sim\rho\nu^{1/2}DU_{0}^{2}t^{1/2}$,
which gives 
\begin{equation}
\widetilde{F}\sim\frac{1}{\Rey^{1/2}}\tau^{1/2}.\label{eq:ViscousScaling4}
\end{equation}

Equations~\eqref{eq:ViscousScaling4}
predicts that the exponent of the early-time scaling, $\beta$, stays
at $1/2$, whereas the coefficient of the scaling, $\alpha$, increases
with decreasing $\Rey$, qualitatively agreeing with our experiments
at intermediate $\Rey$ when $0.7<\Rey<200$ (figure~\ref{fig3}). Quantitatively, we fit
$(\alpha-\alpha^{\infty})$ as a function of $\Rey$ from our experiments using 
\begin{equation}
\alpha(\Rey)-\alpha^{\infty}=\frac{c_{0}}{\Rey^{\gamma}},\label{eq:coefficient1}
\end{equation}
where $\alpha^{\infty}=3 \sqrt{6} /2$ is the asymptotic coefficient at
the high-$\Rey$ limit from the initial-impact self-similar theory in $\S$\ref{subsec:Experimental-evidence}. Our experiments show
$\gamma=0.45\pm 0.4$, consistent with the $\Rey^{-1/2}$ scaling of
\eqref{eq:ViscousScaling4} (the inset of figure~\ref{fig3}\textit{a}).
In addition, we obtain $c_{0}=4.36\pm 0.50$.

Although the simple scaling of \eqref{eq:ViscousScaling4} successfully
explains the early-time scaling of the impact force of viscous drops,
the usage of $\delta_{\nu}$ as the characteristic length scale in
our argument needs a formal justification. Moreover, the simple scaling
only provides the viscous contribution of the impact force. It is
not clear how the viscous impact force couples with the inertial impact
force at finite $\Rey$. When fitting experiments using \eqref{eq:coefficient1},
we simply assume the two forces are additive. This simple assumption
also needs to be justified. Lastly, it is certainly relevant to analytically
calculate the coefficient $c_{0}$ in the scaling \eqref{eq:coefficient1}.

Here, we develop an asymptotic perturbation method to calculate the
impact force of viscous drops at finite $\Rey$ during initial impact
\citep{Orszag}. The starting point of our calculation is the leading-order self-similar dimensionless
radial velocity field inside the boundary layer.
The field was obtained by Philippi and coworkers in analogy to the
shock-induced boundary layers \citep{Philippi:2016bg}, which compares
well with the numerical result: 
\begin{equation}
u_{r}^{\left(0\right)}=\frac{2r}{\upi\sqrt{\delta^{2}\tau-r^{2}}}f'\left(\eta\equiv\frac{\delta}{2}\sqrt{\frac{\Rey}{\delta^{2}\tau-r^{2}}}z\right),\label{eq:radialvelocity}
\end{equation}
where $\delta=\sqrt{6}/2$ is a constant, indicating the spreading contact line  $r_{0}=\delta\sqrt{\tau}$ (figure~\ref{fig1}\textit{a}).\footnote{Notice that we define dimensionless quantities based on the diameter
of liquid drops, instead of the radius of liquid drops used in \citet{Philippi:2016bg},
which modifies the constant coefficients in \eqref{eq:radialvelocity}.} The profile $f'$ is the $\mathrm{erf}$ function and $\eta$ is
introduced as the dimensionless inner variable of the boundary layer.
We assume a perturbation expansion for the inner velocity field, $\left(u_{r},u_{z}\right)$,
in terms of the small parameter $\epsilon=\Rey^{-1/2}$. Thus, the
radial velocity field can be expanded as $u_{r}=u_{r}^{\left(0\right)}+\epsilon u_{r}^{\left(1\right)}+\epsilon^{2}u_{r}^{\left(2\right)}+\mathcal{O}\left(\epsilon^{3}\right)$
and the vertical velocity field as $u_{z}=u_{z}^{\left(0\right)}+\epsilon u_{z}^{\left(1\right)}+\epsilon^{2}u_{z}^{\left(2\right)}+\mathcal{O}\left(\epsilon^{3}\right)$.
From \eqref{eq:radialvelocity} and the mass conservation, we immediately
have 
\[
u_{z}^{\left(0\right)}=0\quad\text{and}\quad u_{z}^{\left(1\right)}=-\frac{4}{\upi\delta}\left[2f+\frac{r^{2}}{\delta^{2}\tau-r^{2}}\eta f'\right].
\]
Likewise, we also expand the dimensionless outer velocity field ($\eta\gg1$),
$\left(U_{r},U_{z}\right)$, in terms of $\epsilon$. The asymptotic
matching condition at the order $\epsilon$ for the vertical velocity
reads \citep{van1975perturbation} 
\[
\epsilon U_{z}^{\left(1\right)}\left(z=0\right)=\lim_{\eta\rightarrow\infty}\epsilon u_{z}^{\left(1\right)}-\lim_{z\rightarrow0}U_{z}^{\left(0\right)}+\mathcal{O}\left(\epsilon^{2}\right).
\]
Using 
\[
\lim_{z\rightarrow0}U_{z}^{\left(0\right)}=-\frac{2z}{\pi\sqrt{\delta^{2}\tau-r^{2}}}\left(2+\frac{r^{2}}{\delta^{2}\tau-r^{2}}\right),
\]
obtained from the mass conservation in the outer flow at $z\rightarrow0$
and expressing $z$ in terms of $\eta$, we obtain 
\[
U_{z}^{\left(1\right)}\left(z=0\right)=-\frac{4}{\upi\delta}\lim_{\eta\rightarrow\infty}\left[2\left(f-\eta\right)+\frac{r^{2}}{\delta^{2}\tau-r^{2}}\eta\left(f'-1\right)\right]+\mathcal{O}\left(\epsilon^{2}\right).
\]
Since $f\left(\eta\right)=\eta-1/\sqrt{\upi}+\mathcal{O}\left(\mathrm{\eta^{-2}e}^{-\eta^{2}}\right)$,
we find that, at the first order of $\epsilon$, the correction of
the vertical velocity of the outer flow at $z=0$ is 
\[
U_{z}^{\left(1\right)}\left(z=0\right)=\frac{8\sqrt{6}}{3\upi^{3/2}}.
\]
Remarkably, the presence of the self-similar boundary layer at finite
$\Rey$ induces at $\mathcal{O}\left(\epsilon\right)$ a uniform velocity in the outer
flow near $z=0$.

\begin{figure}
\begin{centering}
\includegraphics{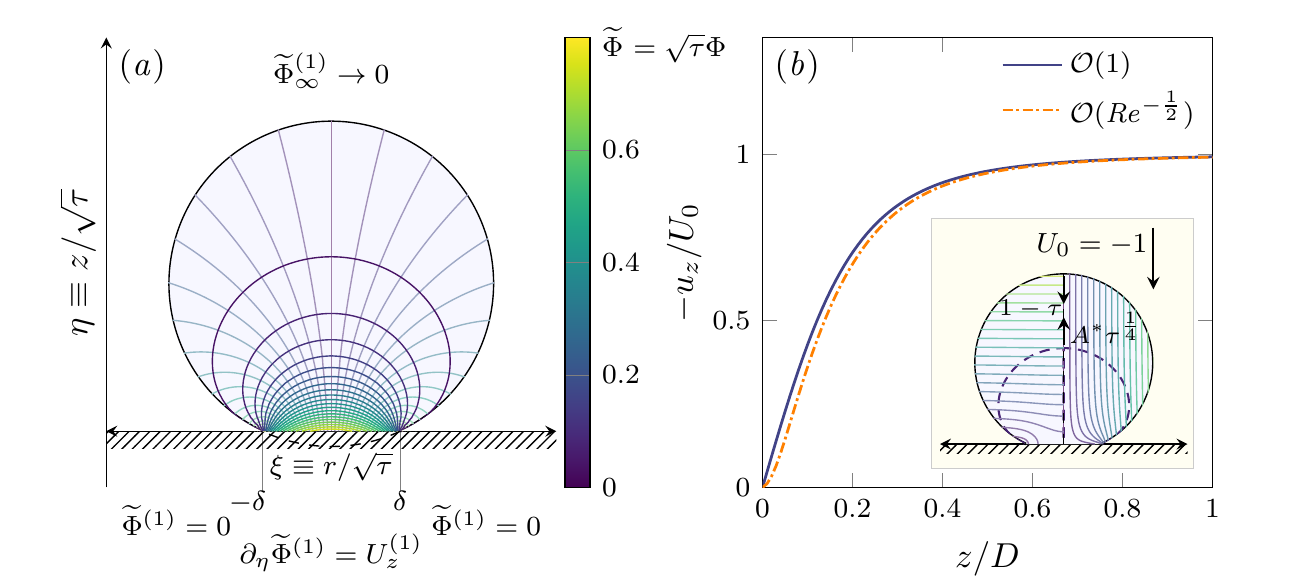} 
\par\end{centering}
\caption{Flow induced by viscous boundary layer. (\textit{a}) Velocity potential
lines (thick) and streamlines (thin) in the self-similar frame of
reference \citep{Philippi:2016bg}, obtained by calculating the self-similar
potential $\widetilde{\Phi}^{(1)}\equiv\sqrt{\tau}\Phi^{(1)}$ from
\eqref{eq:math1}-\eqref{eq:math3}. The potential satisfies Laplace's
equation and the boundary conditions similar to those at $\mathcal{O}\left(1\right)$
(cf. figure 8 in \citet{Philippi:2016bg}). (b) Dimensionless vertical
velocity profile at $r=0$ in the laboratory frame with $\Rey=100$ and
$\tau=0.1$. The thick solid line represents the $\mathcal{O}\left(1\right)$
velocity profile from \citet{Philippi:2016bg}. The dash-dotted line
is our uniformly asymptotic correction after introducing the boundary layer. Notice that the dash-dotted
line is on the right of the solid line, indicating a faster expansion
of the self-similar field in the presence of the boundary layer. The inset
shows a snapshot of the potential lines (left) and streamlines (right)
in the laboratory frame. The dashed line represents the shape of an isobar far from the impact point. While the apex of the drop
travels downward unperturbed as $(1-\tau)$, the isobar propagates
upward as $A^{*}\tau^{1/4}$ in the lab frame (see the text). \label{fig5}}
\end{figure}

With the boundary conditions corrected due to the boundary layer,
the outer velocity field at $\mathcal{O}\left(\epsilon\right)$ are given by an inviscid
problem that can be solved using a potential velocity field $\Phi^{\left(1\right)}$, which satisfies Laplace's equation,
$\nabla^{2}\Phi^{\left(1\right)}=0$, and the set of boundary conditions:
\begin{align}
\frac{\partial}{\partial z}\Phi^{\left(1\right)}=U_{z}^{\left(1\right)}, & \quad\text{at \ensuremath{z=0}, \ensuremath{r<\delta\sqrt{\tau},}}\label{eq:math1}\\
\Phi^{\left(1\right)}=0, & \quad\text{at \ensuremath{z=0}, \ensuremath{r>\delta\sqrt{\tau}},}\label{eq:math2}\\
\Phi^{\left(1\right)}\rightarrow0, & \quad\text{at \ensuremath{z=\infty}.}\label{eq:math3}
\end{align}
The mathematical structure of the problem is the same as the one solved
at the zeroth order after changing the frame of reference (figure~\ref{fig5}\textit{a}).
Hence, the method used in \citet{Philippi:2016bg} for solving the
solution of the outer flow at the zeroth order can be directly used
to obtain the flow field at $\mathcal{O}\left(\epsilon\right)$.
The uniform asymptotic expansion of the velocity field at $\mathcal{O}\left(\epsilon\right)$
can be obtained by matching $u_{z}=\epsilon u_{z}^{\left(1\right)}$
and $U_{z}=U_{z}^{\left(0\right)}+\epsilon\nabla\Phi^{\left(1\right)}$. 
An example of a uniform asymptotic expansion of the vertical velocity profile at $r=0$ at $\mathcal{O}\left(\epsilon\right)$ is depicted for $\epsilon=0.1$
and $\tau=0.1$ and compared with the profile at $\mathcal{O}\left(1\right)$
in figure~\ref{fig5}(\textit{b}). The smaller vertical velocity at $\mathcal{O}\left(\epsilon\right)$ at a fixed $z$ indicates a faster propagation of
the self-similar field in the presence of the boundary layer. In other words, the boundary layer affects the self-similar pressure field, making it propagate
faster than that in the inviscid case at the high $\Rey$ limit (figure~\ref{fig5}\textit{b}).

Conveniently, many results at leading order can be renormalised to obtain results 
at the order of $\mathcal{O}\left(\epsilon\right)$ by simply replacing the
impact velocity $U_{0}$ with $U_{0}\left(1+\epsilon U_{z}^{\left(1\right)}\right)$.
It is straightforward to show that in comparison with the
impact force at the zeroth order \eqref{eq:Force-1}, the dimensionless
force at the first order near $\tau=0^{+}$ is 
\begin{equation}
\widetilde{F}=\frac{3\sqrt{6}}{2}\left(1+\frac{8\sqrt{6}}{3\upi^{3/2}}\frac{1}{\Rey^{1/2}}\right)\tau^{1/2}+\mathcal{O}\left(\Rey^{-1}\right),
\end{equation}
which gives the coefficient and the exponent of the early-time scaling
in $\widetilde{F}=\alpha\tau^{\beta}$ 
\begin{equation}
\alpha=\frac{3\sqrt{6}}{2}\left(1+\frac{8\sqrt{6}}{3\upi^{3/2}}\frac{1}{\Rey^{1/2}}\right)=\alpha^{\infty}+\frac{24}{\upi^{3/2}}\frac{1}{\Rey^{1/2}}\quad\text{and}\quad\beta=1/2,\label{eq:viscous_alpha}
\end{equation}
where $\alpha^{\infty}=3\sqrt{6}/2$ is the asymptotic value of $\alpha$
when $\Rey\to\infty$ predicted by the initial-impact self-similar theory \eqref{eq:DimensionlessForce-1}.
Equation~\eqref{eq:viscous_alpha} directly confirms the $\Rey^{-1/2}$
scaling for $\alpha$ at finite $\Rey$ and, therefore, verifies the usage of the boundary
layer thickness $\delta_{\nu}$ as the relevant length scale in the
simple scaling argument. Second, it shows that the inertial and viscous
impact forces are additive as shown in \eqref{eq:coefficient1}. Third, it gives $c_{0}=24/(\upi^{3/2})\approx4.31$,
quantitatively agreeing with our experiments $c_{0}=4.36\pm0.50$. As such, equation \eqref{eq:viscous_alpha}
quantitatively describes the experimental trends of $\alpha(\Rey)$
and $\beta(\Rey)$ without fitting parameters (the dashed-dotted lines
in figure~\ref{fig3}).

The simple picture that the viscous boundary layer effectively increases
the propagation speed of the self-similar pressure field also allows us to quantitatively
predict the trends of $t_{\max}$ and $F_{\max}$ as a function of
$\Rey$. To determine $t_{\max}$, we first analyze the propagation of isobars far away from the impact point within an impacting drop. We find that the isobars travel as $\left(U_{0}D^3t\right)^{1/4}$ at the high $\Rey$ limit (appendix~\ref{appInertial}). Notice that the propagation speed of isobars is different from the length scale of the self-similar structure. The former indicates the location of constant-pressure contours, whereas the latter arises from the self-similar arguments when constructing the self-similar pressure field (appendix~\ref{appInertial}). When the
isobar of a preset high pressure touches the upper surface of the liquid drop, which
moves downward ballistically as $\left(D-U_{0}t\right)$, the initial-impact regime terminates.
Hence, $t_{\max}$, the boundary of the initial-impact regime, can
be estimated simply from $A\left(U_{0}D^{3}t_{\max}\right)^{1/4}=\left(D-U_{0}t_{\max}\right)$,
where $A$ is a geometric factor that accounts for the threshold at
which the apex starts to be affected by the self-similar pressure
field. In the dimensionless form, the condition
simply writes as $A\tau_{\max}^{1/4}=\left(1-\tau_{\max}\right)$ (see the schematic in figure~\ref{fig5}\textit{b}).
From the asymptotic value of $\tau_{\max}$ at the high-$\Rey$ limit,
$\tau_{\max}^{\infty}\approx0.18\pm0.05$, we find $A=1.24\pm0.10$,
on the order of one as expected. At finite $\Rey$, we assume that
the non-monotonic trend of the impact forces also arises from the
termination of the initial impact. Nevertheless, the propagation of
the isobar should be corrected due to the presence of the
boundary layer at finite $\Rey$. The isobar now propagates as $A^{*}\tau_{\max}^{1/4}$ with a renormalised $A^{*}=A\sqrt{1+\epsilon U_{z}^{\left(1\right)}}$.
The peak time is then given by the solution of the polynomial
\begin{equation}
\left(1-\tau_{\max}\right)^{4}-A^{4}\left(1+U_{z}^{\left(1\right)}\Rey^{-1/2}\right)^{2}\tau_{\max}=0.\label{eq:tmax1}
\end{equation}
Notice that the ballistic motion of the apex of the drop is not affected
by the correction $\epsilon U_{z}^{\left(1\right)}$, since the upper surface of the drop
has not experienced the impact during the initial impact and, therefore,
is not influenced by the impact-induced boundary layer. Equation~\eqref{eq:tmax1}
successfully predicts the decrease of $\tau_{\max}$ with decreasing
$\Rey$, which quantitatively matches $\tau_{\max}(\Rey)$ in three
decades of $\Rey$ (the dash-dotted line in figure~\ref{fig4}\textit{a}).

\begin{figure}
\begin{centering}
\includegraphics{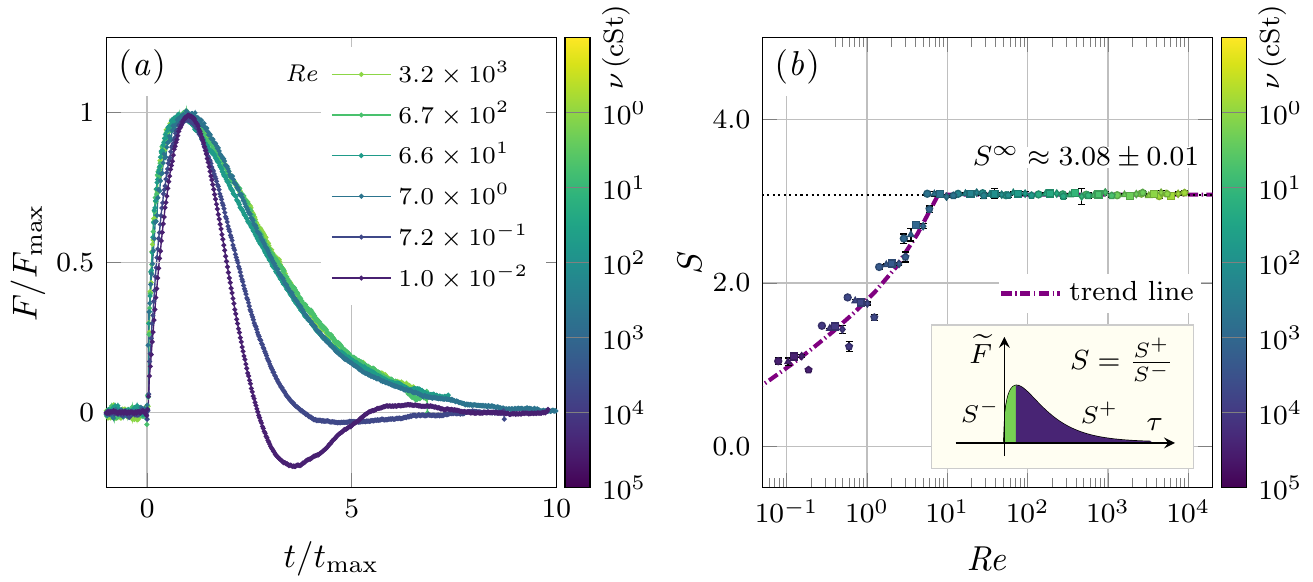} 
\par\end{centering}
\caption{Shape of impact forces. (\textit{a}) Rescaled impact forces of the six impacts shown in figure~\ref{fig2}(\textit{a}).
$F$ is normalised by the maximum impact force, $F_{\max}$, and $t$
is normalised by the peak time, $t_{\max}$. The rescaled impact forces
collapse into a master curve for $\Rey>7$. Notice that the curves
are the same if plotting in terms of dimensionless quantities $\widetilde{F}/\widetilde{F}_{\max}$
versus $\tau/\tau_{\max}$. (\textit{b}) Symmetry factor, $S$, defined
as the quotient of the time integral of $\widetilde{F}$ before and
after $\tau_{\max}$, as a function $\Rey$ (see inset). The horizontal
dashed line indicates the asymptotic value at high $\Rey$. The thick
dash-dotted line is a fitting as a guide of eyes.\label{fig6}}
\end{figure}

Although $F_{\max}$ has been extensively investigated and the scaling
of $F_{\max}$ with the inertial force $\rho D^{2}U_{0}^{2}$ has
been reported in several previous experiments \citep{Grinspan:2010gt,Li:2014gq,Soto:2014fx,BinZhang:2017ix},
to the best of our knowledge, a quantitative description of $F_{\max}$
as a function of $\Rey$ is still not available. Here, we propose
a simple model for $F_{\max}(\Rey)$. Our calculation is based on
an interesting observation: the overall shape of the rescaled impact
force $F/F_{\max}$ is invariant when plotted against the rescaled
time $t/t_{\max}$ in the regime of high and intermediate $\Rey$.
From high to intermediate $\Rey$, $F(t)$ is highly asymmetric with
respect to $t_{\max}$ (figures~\ref{fig1}\textit{b} and \ref{fig2}\textit{a}):
the increase of the impact force is fast before $t_{\max}$ and decays
much slower after $t_{\max}$. In contrast, for low-$\Rey$ impacts, $F(t)$
becomes more symmetric (figure~\ref{fig2}\textit{a}). The rise and
decay of $F(t)$ show a similar time scale. To quantify the change
of the shape of $F(t)$, we define a symmetry factor, $S\equiv\int_{0}^{t_{\mathrm{max}}}F\left(t\right)\,\mathrm{d}t/\int_{t_{\mathrm{max}}}^{\infty}F\left(t\right)\,\mathrm{d}t$,\footnote{Notice that for the impact force of very low $\Rey$, $F(t)$ oscillates
at long times and exhibits negative impact pressures (figure~\ref{fig2}\textit{a}; for explanation see $\S$\ref{subsec:viscoelasticity}). In this
case, we replace the upper limit of the integral in the denominator
$t=\infty$ to a finite $t_{0}$, the time when $F(t)$ first crosses
zero.} which is shown as a function of $\Rey$ in figure~\ref{fig6}(\textit{b}).
Interestingly, $S$ reaches a plateau $S^{\infty}=3.08\pm0.01$ when
$\Rey>7$, showing that the impulse of impacts before $t_{\max}$ invariably annihilates a quarter of the total momentum of liquid drops
irregardless $\Rey$ as long as $\Rey>7$. The constant plateau of
$S$ suggests that the shapes of the rescaled impact force, $F(t/t_{\max})/F_{\max}$, are invariant with changing $\Rey$ and can be collapsed into a master
curve when $\Rey>7$. We directly confirmed this hypothesis in our
experiments (figure~\ref{fig6}\textit{a}). The collapse of $\widetilde{F}(\tau)$ at high $\Rey$ without the rescaling $\widetilde{F}/\widetilde{F}_{\max}$ and $\tau/\tau_{\max}$
has also been reported in a recent experiment, where
$\widetilde{F}_{\max}=\widetilde{F}_{\max}^{\infty}$ and $\tau_{\max}=\tau_{\max}^{\infty}$
are constant (cf. figure~6 in \citet{BinZhang:2017ix}). Finally,
since the integral of the force is equal to the momentum of the drop,
\[
\int_{0}^{\infty}\widetilde{F}\left(\tau\right)\,\mathrm{d}\tau=\frac{\pi}{6},
\]
it is straightforward to show that 
\begin{equation}
\widetilde{F}_{\mathrm{max}}=\widetilde{F}_{\mathrm{max}}^{\infty}\frac{\tau_{\mathrm{max}}^{\infty}}{\tau_{\mathrm{max}}}.\label{eq:fmax}
\end{equation}
Using \eqref{eq:tmax1} and the asymptotic value of $\widetilde{F}_{\max}$
at high $\Rey$, $\widetilde{F}_{\max}^{\infty}\approx0.83$, \eqref{eq:fmax}
quantitatively predicts the trend of $\widetilde{F}_{\max}(\Rey)$
for over five decades of $\Rey$ between $0.3$ and $10^{4}$
(the dash-dotted line in figure~\ref{fig4}\textit{b}).

\subsubsection{The effect of viscoelasticity on impact forces \label{subsec:viscoelasticity}}

At even lower $\Rey$ below 0.7, $\beta$ increases above $1/2$ and
deviates from the scaling predicted for impact forces near $\tau=0^{+}$
at finite $\Rey$. The data also show a strong scatter in this regime
(figure~\ref{fig3}\textit{b} and figure~\ref{fig4}\textit{a}).
To experimentally achieve these low-$\Rey$ impacts, we had to use
silicone oils of high viscosities above $10\,000\;\mathrm{cSt}$. Silicone
oils of such high viscosities are made of polymerised siloxanes of
high molecular weights, which exhibit obvious viscoelasticity during fast impacts. The increase of $\beta$ can be attributed
to the increase of elasticity. In the elastic limit, the kinetic energy
of an impinging drop is converted into the elastic potential of the
deformed drop. The deformation of the elastic drop can still be approximated
as $d\approx U_{0}t$ at short times. The elastic strain in the deformed
drop is $d/r_{0}$ with $r_{0}\approx\sqrt{dD}$ and the volume of
the deformed region is $\sim r_{0}^{2}d$. The energy balance
in the elastic limit can then be written as 
\begin{equation}
Fd\sim E\frac{d}{r_{0}}r_{0}^{2}d,\label{eq:ElasticScaling}
\end{equation}
which gives 
\begin{equation}
\widetilde{F}\sim\frac{E}{\rho U_{0}^{2}}\tau^{3/2}\label{eq:DimensionlessElasticScaling}
\end{equation}
in the dimensionless form, where $E$ is the elastic modulus of the drop.
The $\tau^{3/2}$ scaling is the well-known result for the
impact force of elastic spheres with Hertzian contacts. A detailed calculation shows 
\begin{equation}
\widetilde{F}=\frac{2\sqrt{2}}{3}\frac{E}{\rho U_{0}^{2}}\tau^{3/2}\label{eq:herziandimensionless}
\end{equation}
(see appendix~\ref{appN}). The large exponent 3/2 in the pure elastic
limit qualitatively explains the increase of $\beta$ as the elastic
effect of high-molecular-weight silicone oils gradually sets in. In
the presence of viscoelasticity, $\Rey$ is no longer a proper dimensionless
number for scaling the data, which results in the strong scatter of
data shown in figures~\ref{fig3}(\textit{b})~and~\ref{fig4}(\textit{a}).

The effect of viscoelasticity of high-molecular-weight silicone oils
can also be seen from the overall shape of impact forces (figures~\ref{fig2}\textit{a}~
and~\ref{fig6}\textit{a}). While $F(t)$ of low-viscosity silicone
oils are highly asymmetric, $F(t)$ for high-molecular-weight silicone
oils becomes more symmetric with respect to $t_{\max}$, approaching
the symmetric impact force of elastic spheres. Quantitatively, the
symmetry factor, $S$, of high-$\Rey$ impacts is large with $S^{\infty}=3.08\pm0.01$
(figure~\ref{fig6}\textit{b}). In contrast, $S$ of the high-viscosity
silicone oils gradually approaches 1, signalling a perfect symmetric
curve similar to the impact force of elastic spheres. The elastic
effect becomes even more obvious for silicone oils of very high viscosity
above $30\,000\;\mathrm{cSt}$. The drops made of these oils bounce
upward slightly towards the end of impact processes due to their elasticity
(see the last column of the Supplementary Video), leading to negative
impact pressures and oscillating impact forces (figure~\ref{fig2}\textit{a}).

To conclude $\S$\ref{subsec:initialimpact}, we summarise the early-time
scaling of impact forces during initial impact at different regimes
in table~\ref{tab:kd}.

\begin{table}
\centering{}%
\begin{tabular}{cccc}
Impact regime  & $\Rey$  & $\alpha$  & $\beta$ \tabularnewline
\midrule 
Inertial  & $>200$  & $\frac{3\sqrt{6}}{2}$  & $1/2$\tabularnewline
Viscous  & $0.7-200$  & $\frac{3\sqrt{6}}{2}\left(1+\frac{8\sqrt{6}}{3\upi^{3/2}}\frac{1}{\Rey^{1/2}}\right)$  & $1/2$\tabularnewline
Elastic  & N/A  & $\frac{2\sqrt{2}}{3}\frac{E}{\rho U_{0}^{2}}$  & $3/2$\tabularnewline
\end{tabular}\caption{Early-time scaling of impact forces, $\widetilde{F}=\alpha\tau^{\beta}$,
near $\tau=0^{+}$ at three different impact regimes\label{tab:kd}}
\end{table}

\subsection{Inertia-driven drop spreading}

In this section, we will investigate the dynamics of drop impacts
during spreading after $t_{\max}$. We shall limit our discussion
to the inertia-driven high-$\Rey$ impacts. After the self-similar field
expands across the drop, the impact force decreases and the apex of
the drop decelerates visibly (figure~\ref{fig1}). The strong self-similar pressure
gradients diminish. Driven by inertia alone,
the drop enters into the spreading regime \citep{Eggers:2010gd}.
The spreading is eventually checked by either viscous or capillary
forces, which dictates the maximum spreading diameter of the drop
at the end of the spreading regime \citep{Yarin:2006al}. In the case
when We $\gg1$, the spreading is stopped by the inertia-viscous balance.
The upper limit of the spreading regime can thus be estimated as $t_{b}=h_{\max}(t_{b})^{2}/\nu$,
which balances the boundary layer $\delta_{\nu}\sim\sqrt{\nu t}$
with the height of the drop $h_{\max}(t)$ \citep{Roisman:2009cg}. For low-viscosity liquids,
this regime of \emph{inertial spreading} spanning between $t_{\max}$
and $t_{b}$ dominates the behaviour of the impacting drop
(figure~\ref{fig1}\textit{a}).

An exact solution for the inertial spreading is still not available.
Eggers and co-workers proposed a self-similar solution, which is exact
at the asymptotic limit when $t\rightarrow\infty$ \citep{Eggers:2010gd}.
Although this theory successfully predicts the asymptotic self-similar
scaling of the shape of spreading drops \citep{Lagubeau:2012ba},
it does not provide a full description for the dynamics of the spreading
drop at finite times. Inspired by the asymptotic self-similar solution,
we show here a closed-form exact solution for inertial spreading at
finite times.

Since inertia dominates the spreading process, the dynamics of the
drop follow the continuity and Euler equations in dimensionless cylindrical
coordinates: 
\begin{eqnarray}
\bnabla\cdot\left(r\mathbf{u}\right) & = & 0,\label{eq:cont}\\
\partial_{\tau}\mathbf{u}+\left(\mathbf{u}\cdot\bnabla\right)\mathbf{u} & = & -\bnabla p,\label{eq:Euler}
\end{eqnarray}
where $\mathbf{u}\equiv\left(u_{r},u_{z}\right)$ is the dimensionless
axisymmetric velocity field, $p$ is the dimensionless pressure and
$\bnabla\equiv\left(\partial_{r},\partial_{z}\right)$. The problem
is closed with the rigid-wall boundary condition at the impacted surface
$z=0$, and the kinematic and dynamic conditions at the interface
$z=h\left(r,\tau\right)$, 
\begin{eqnarray}
\left.u_{z}\right|_{z=0} & = & 0,\label{eq:bottom_bc}\\
\left.\partial_{t}h+u_{r}\partial_{r}h-u_{z}\right|_{z=h\left(r,\tau\right)} & = & 0,\label{eq:top_kc}\\
\left.p\right|_{z=h\left(r,\tau\right)} & = & 0.\label{eq:top_dc}
\end{eqnarray}
Since We $\gg1$, $p$ is a constant at the interface, which we can
set to zero.

We generalise the self-similar hyperbolic velocity field proposed
by \citet{Eggers:2010gd} into $\mathbf{u}\left(r,\tau\right)=\left(f\left(\tau\right)r,-2f\left(\tau\right)z\right)$,
which automatically satisfies equations \eqref{eq:cont} and \eqref{eq:bottom_bc}.
The shape of the drop can be generally written as $h\left(r,\tau\right)=\omega H\left(\zeta\equiv\omega^{1/2}r\right)$,
where $\omega\equiv\omega\left(\tau\right)$ is an unknown function.
Note that the form of $h\left(r,\tau\right)$ conserves the volume,
a crucial ingredient of the spreading regime. Replacing the \emph{ansatz}
for $\mathbf{u}$ and $h$ in the kinematic condition \eqref{eq:top_kc},
and expressing the equation in terms of the self-similar variable
$\zeta$, we obtain 
\begin{equation}
\left(\omega'+2\omega f\right)\left(H+\frac{1}{2}\zeta H'\right)=0,
\end{equation}
which is satisfied for any $H$ when $f=-\omega'/\left(2\omega\right)$.
Replacing this value of $f\left(\tau\right)$ in \eqref{eq:Euler},
we find the pressure 
\begin{equation}
p\left(r,z,\tau\right)=-\frac{1}{2}\omega^{-1/2}\left(\omega^{-1/2}\right)''\zeta^{2}-\frac{1}{2}\frac{\omega''}{\omega}z^{2}+\Omega\left(\tau\right),\label{eq:pressure}
\end{equation}
where $\Omega\left(\tau\right)$ is an arbitrary function of time.
Finally, the dynamic condition \eqref{eq:top_dc} enforces $p=0$
at $z=h(r,\tau)$, which leads to an algebraic equation for $H$ 
\begin{equation}
H\left(\zeta\right)=\sqrt{\left(\frac{2\Omega\left(\tau\right)}{\omega\omega''}\right)-\left(\frac{\omega^{-1/2}\left(\omega^{-1/2}\right)''}{\omega\omega''}\right)\zeta^{2}}.
\end{equation}
Since $H\left(\zeta\right)$ is strictly a function of $\zeta$ and
not $\tau$, the quantities in parentheses have to be constants. Defining
those constants as $H_{0}^{2}$ and $R^{-3}H_{0}^{3}/2$ respectively
and introducing $\widehat{\omega}\equiv R^{-1}H_{0}\omega$, we have
\begin{equation}
\frac{2\Omega\left(\tau\right)}{\widehat{\omega}\widehat{\omega}''}=R^{2},\quad\frac{\widehat{\omega}^{-1/2}\left(\widehat{\omega}^{-1/2}\right)''}{\widehat{\omega}\widehat{\omega}''}=\frac{1}{2}.
\end{equation}
While the first equation is trivial, the solution of the second equation
is given by a closed form 
\begin{equation}
t\left(\widehat{\omega}\right)=t_{0}+t_{1}T\left(\widehat{\omega}\right),\label{eq:omega}
\end{equation}
where $T\left(x\right)\equiv2x^{-1/2}\left(1+x^{3}\right)^{1/2}-3\left(1+x^{3}\right)^{1/3}F_{\left[-\frac{1}{3},\frac{1}{6},\frac{2}{3}\right]}\left(\left[1+x^{3}\right]^{-1}\right)$
and $F_{\left[m,n,p\right]}(x)$ is a hypergeometric function. Thus,
an exact self-similar solution of the Euler equations is obtained.
The solution also satisfies the full Navier-Stokes equations in the
bulk without boundary conditions. $t_{0}$ and $t_{1}$ are integral
constants, which can be fixed by requesting that the position and
the speed of the drop apex at $\tau=0$ are $h=1$ and $u_{z}=-1$, respectively.
These requirements lead to $t_{1}=R/\sqrt{1+R^{3}}$ and $t_{0}=-t_{1}T\left(R^{-1}\right)$.
$R$ is finally obtained by setting the volume of the drop at its
dimensionless value $\upi/6$, which yields $R=1/2$, i.e., the dimensionless
radius of the drop.

\begin{figure}
\begin{centering}
\includegraphics{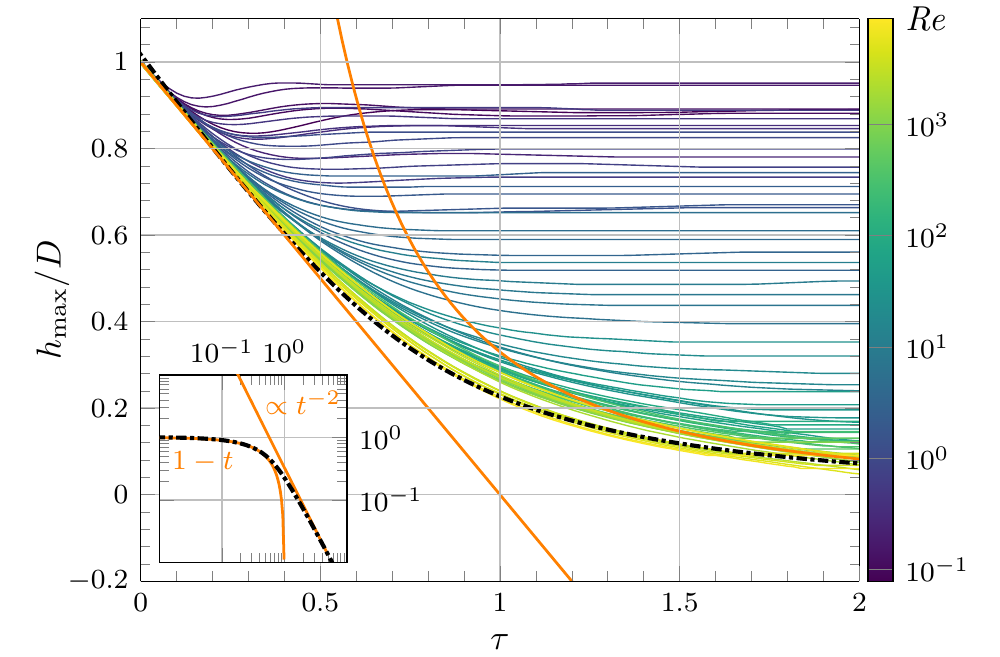} 
\par\end{centering}
\caption{Temporal variation of the apex of liquid drops, $h_{\max}(\tau)$.
Solid thin curves are from experiments at different $\Rey$ with the
value of $\Rey$ indicated by the colour bar on the right. The black
dash-dotted line is the prediction of our exact solution \eqref{eq:hmax}.
The solid orange line on the left has a slope of $-1$, indicating
the linear descent of liquid drops with impact velocity $U_{0}$.
The solid orange line on the right indicates a power-law scaling $\tau^{-2}$,
i.e., the asymptotic behaviour when $\tau\rightarrow\infty$ predicted
by previous studies \citep{Eggers:2010gd,Roisman:2009ee,Lagubeau:2012ba}.
The inset shows the asymptotic behaviour of our solution in the limit of $\tau\rightarrow0$ and $\tau\rightarrow\infty$ in
a log-log scale.\label{fig7}}
\end{figure}

Particularly, with $\widehat{\omega}$ from \eqref{eq:omega}, we
have the shape of the spreading drop 
\begin{equation}
h\left(r,\tau\right)=\widehat{\omega}\sqrt{R^{2}-\frac{1}{2}\widehat{\omega}r^{2}},\label{eq:interface}
\end{equation}
which gives the height of the drop  
\begin{equation}
h_{\max}\equiv h\left(r=0,\tau\right)=R\widehat{\omega}. \label{eq:hmax}
\end{equation}
$h_{\max}(\tau = 0^+)$ naturally captures the linear
descent of the apex with constant velocity $U_{0}$ ($-1$ in dimensionless
units) as imposed by the boundary condition (figure~\ref{fig7}).
More importantly, it provides the correct asymptotic limit of inertia-driven
spreading when $\tau\rightarrow\infty$. Since the series expansion of $T\left(x\right)$ at $x=0$
is given by $T\left(x\right)=2x^{-1/2}-T_{0}+\mathcal{\mathcal{O}}\left(x^{5/2}\right)$,
where $T_{0}=3^{3/2}\Gamma^{3}\left(\frac{2}{3}\right)/\left(2^{2/3}\upi\right)$, the asymptotic behaviour of $h_{\mathrm{\max}}$ as $\tau\rightarrow\infty$ is 
\[
\lim_{\tau\rightarrow\infty}h_{\max}\left(\tau\right)=\left(\frac{4R^{3}}{1+R^{3}}\right)\left(\tau+\tau_{\infty}\right)^{-2},
\]
where $\tau_{\infty}=t_{1}\left(T\left(R^{-1}\right)+T_{0}\right)$.
Numerically, 
\begin{equation}
\lim_{\tau\rightarrow\infty}h_{\max}(\tau)\approx0.44\left(\tau+0.31\right)^{-2}\sim\tau^{-2}.\label{eq:asympototiclimit}
\end{equation} 
In comparison, \citet{Roisman:2009ee} showed in numerical simulations $h_{\max}(\tau)=0.39\left(\tau+0.25\right)^{-2}$
when $\tau\to\infty$. \citet{Lagubeau:2012ba} showed in experiments
$h_{\max}(\tau)=0.49\left(\tau+0.43\right)^{-2}$ when $\tau\to\infty$.
The asymptotic limit of our self-similar solution \eqref{eq:asympototiclimit}
quantitatively matches these observations. Finally, we directly compare our experimentally measured height of the drop, $h_{\max}/D$
with \eqref{eq:hmax} (figure~\ref{fig7}). The theory shows
a quantitative agreement with experiments at high $\Rey$ over the entire range of $\tau$ without fitting parameters.

\begin{figure}
\begin{centering}
\includegraphics{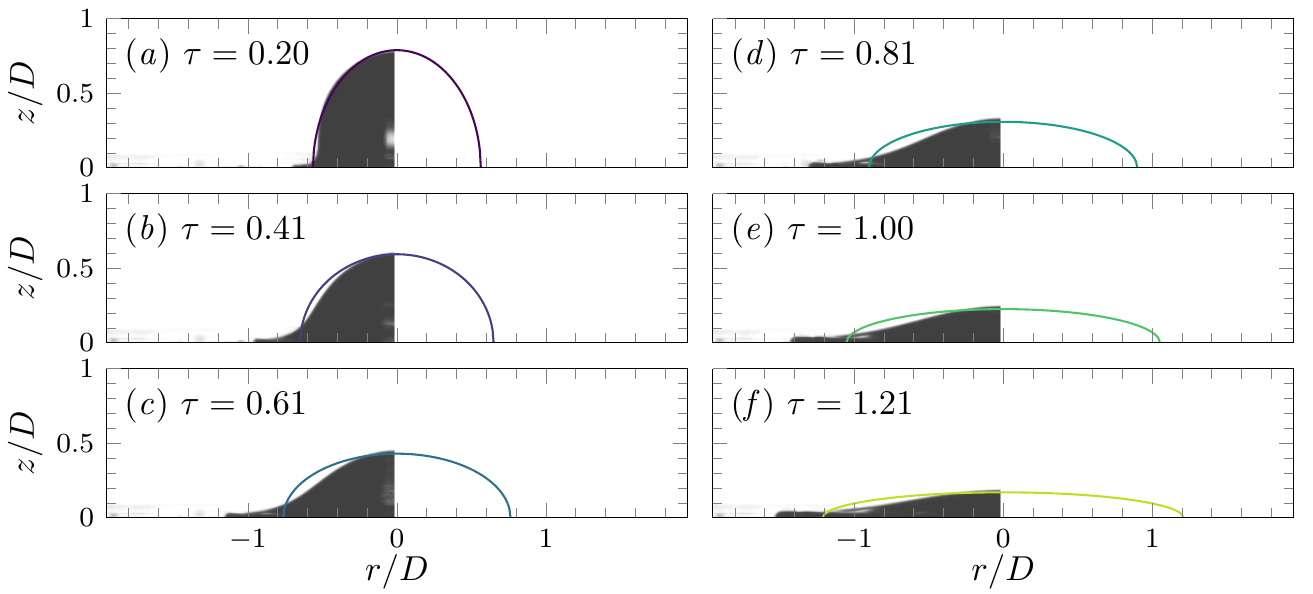} 
\par\end{centering}
\caption{Temporal evolution of the shape of a spreading drop. The figures (\textit{a-f})
show snapshots of the drop at six different times during impact. Solid
lines are our self-similar solution \eqref{eq:interface}. Grey pictures are experimental images taken by high-speed
photography. Thanks to the axial symmetry, we show only the left half
of the experimental images for clarity. The viscosity, impact velocity
and $\Rey$ of the impacting drop are $2\,\mathrm{cSt}$, $1.55\;\mathrm{m/s}$
and $1599$, respectively. The dimensionless times $\tau$ are indicated
in the plots. \label{fig8}}
\end{figure}

Although the self-similar solution quantitatively predicts the height
of spreading drops, the limitation of the solution is obvious. First,
since the Euler equations and the boundary conditions apply only outside
the boundary layer, the solution fails to describe the dynamics of
the contact line at the air-liquid-solid interface (figure~\ref{fig8}).
Hence, the exact solution cannot quantitatively predict the dynamics
of the spreading lamella \citep{Eggers:2010gd}. Second,
the solution also fails to quantitatively capture the decay of the
impact force in the spreading regime. We calculate the impact force
by integrating the pressure at $z=0$ from \eqref{eq:pressure} over
the contact area. Although the calculated force shows a non-monotonic
trend, the numerical value fits the experimental result only at long
times when $\tau\gtrsim1$ (figure~\ref{fig1}\textit{b}). When the
original asymptotic self-similar solution by Eggers \textit{et al.}
is used, where the shape of the drop at finite times is obtained numerically
by fitting either experimental or numerical results \citep{Eggers:2010gd,Lagubeau:2012ba},
the predicted impact force monotonically decreases with $\tau$ and
shows a better fitting at slightly lower $\tau$ (see appendix~\ref{appA}).

\section{Conclusion and outlook\label{sec:outlook}}

By synchronising force sensory with the high-speed photography, we
simultaneously measured both the kinematics and dynamics of liquid-drop impacts
over a wide range of $\Rey$. Our experiments on the early-time scaling
of impact forces verified that the initial impact of a liquid
drop at high $\Rey$ is governed by upward expanding self-similar pressure and velocity
fields. The expanding speed of the self-similar fields is of the same order of magnitude as the impact speed of the
liquid drop. The prediction of the initial-impact self-similar theory breaks
down when $\Rey\lesssim200$, where viscous dissipation becomes important.
Using a perturbation method, we quantitatively analysed the early-time
scaling of the impact force of viscous drops at finite $\Rey$. Our calculation provided a quantitative description
of the maximum force ($F_{\max}$) and the peak time ($t_{\max}$)
as a function of $\Rey$. Lastly, we also discussed the influence
of viscoelasticity on the temporal evolution of impact forces of high-viscosity
silicone oils. In the spreading regime of drop impacts, we generalised
the asymptotic self-similar solution proposed by Eggers and co-workers \citep{Eggers:2010gd}
and found an exact solution for inertia-driven drop spreading at finite times at high
$\Rey$. Our solution quantitatively predicts the height of spreading
drops. The discrepancy between the exact solution and experiments
on the temporal evolution of contact lines and impact forces
reveals the limit of the self-similar approach in predicting drop-spreading
dynamics. In summary, our systematic experiments illustrate the detailed
temporal evolution of impact forces across inertial, viscous and viscoelastic
regimes. The corresponding theoretical analysis provides a quantitative
understanding of the early-time scaling of impact forces in these
different impact regimes. Finally, our exact self-similar solution
on inertia-driven drop spreading extends the well-known asymptotic
self-similar scaling to finite times and provides a parameter-free
description of the height of spreading drops.

Our work also poses new questions and directions. Theoretically, the
logical next step is to incorporate the exact solution of the Euler
equations with the solution of the boundary layer \citep{Eggers:2010gd}
and quantitatively predict the rim dynamics of liquid lamella \citep{Roisman:2002iv}
and the temporal evolution of impact forces during spreading. More
importantly, a theoretical understanding is needed to bridge the two
self-similar regimes at high $\Rey$, which should illustrate how the
self-similar spreading establishes from the expanding self-similar fields at the end of initial impacts. This transition is particularly important given that
the maximum impact force occurs during the transition. Lastly, it
is also interesting to extend the self-similar solution of drop spreading
at high $\Rey$ into the spreading of viscous drops at finite $\Rey$.
Experimentally, we have showed that high-speed imaging and fast force
measurement are two complementary tools. While high-speed imaging
can accurately resolve the variation of the shape of impacting drops
during spreading, force measurement reveals the unique signature of
drop dynamics during initial impact. Although the use of high-speed
photography has become a routine in the study of drop impacts \citep{Josserand:2016jf},
the combination of the two has not been frequently implemented. A
broader application of the combined techniques will certainly deepen our
understanding of liquid-drop impacts.

\section*{Acknowledgments\label{sec:acknowledgments}}

We thank G. De Hoe, F. Japardi, J. Wang, and W. Teddy for the help with experiments.
The research was supported by NSF CAREER DMR-1452180. L. G. was partially
supported by Conicyt FCHA/Postdoctorado Becas Chile 74160007 and Conicyt
PAI/IAC 79160140.

\appendix

\section{Validation of impact force measurements\label{appN}}
\begin{figure}
	\begin{centering}
		\includegraphics{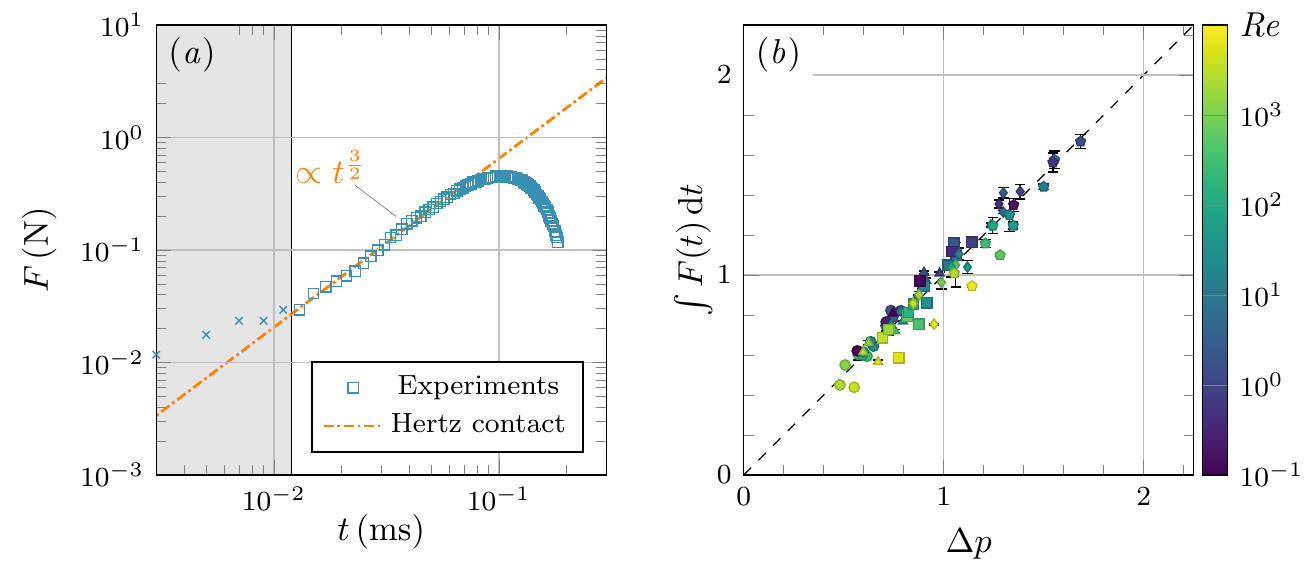} 
		\par\end{centering}
	\caption{Validation of our experimental method. (\textit{a}) Temporal evolution
		of the impact force of a neoprene rubber ball with $D=4.5$ mm and
		$U_{0}=0.4$ m/s. The impact force scales as $t^{3/2}$ near $t=0^{+}$
		as predicted by the Hertz-contact theory \citep{LandauLifshitz:Elasticity}.
		The shaded area indicates the rise time of the force sensor. (\textit{b})
		The impulse of impacts, $J=\int_{0}^{\infty}F\mathrm{d}t$, versus the momentum of liquid drops,
		$\Delta p=mU_{0}$, where $m=\pi\rho D^3/6$
		is the mass of liquid drops. The dashed line indicates $J=\Delta p$
		as requested by momentum conservation. \label{fig9}}
\end{figure}
To verify our experimental method for measuring impact forces, we
have conducted two independent tests. First, we measure the temporal
evolution of the impact force of elastic spheres, a well-known result
in contact mechanisms \citep{LandauLifshitz:Elasticity}. When a non-adhesive
elastic sphere of radius $R$ impacts on an infinite elastic plane,
the impact force is governed by the classical Hertzian contact stress,
\begin{equation}
F=\frac{4}{3}E^{*}R^{1/2}d^{3/2},
\end{equation}
where 
\begin{equation}
\frac{1}{E^{*}}=\frac{1-\nu_{1}^{2}}{E_{1}}+\frac{1-\nu_{2}^{2}}{E_{2}}.
\end{equation}
Here, $E_{1}$ and $E_{2}$ are the elastic moduli and $\nu_{1}$
and $\nu_{2}$ are the Poisson's ratios of the sphere and the plane,
respectively. $d$ is the displacement of the sphere. When $t\to0$,
$d=U_{0}t$. Thus, 
\begin{equation}
\lim_{t\rightarrow0}F=\frac{4}{3}E^{*}R^{1/2}U_{0}^{3/2}t^{3/2},\label{eq:herzian}
\end{equation}
which leads to an early-time scaling
\begin{equation}
\widetilde{F}=\frac{2\sqrt{2}}{3}\frac{E^{*}}{\rho U_{0}^{2}}\tau^{3/2}\label{eq:elasticscaling}
\end{equation}
in the dimensionless form. Our experiments quantitatively agree with
the prediction of \eqref{eq:elasticscaling}. Experimentally, the early-time
scaling of the impact force of elastic spheres shows a power-law scaling $\widetilde{F}=\alpha\tau^{\beta}$
with $\beta=1.49\pm0.04$ and $\alpha=\left(1.45\pm0.75\right)\times10^{5}$
(figure~\ref{fig9}\textit{a}), where the errors are obtained from
five independent runs. In comparison, theoretically, $\beta=3/2$
and $\alpha=7.9\times10^{4}$, where we use the material properties
of neoprene rubber $\rho=1.23$ g/cm$^{3}$, $E_{1}=12.33$ MPa and
$\nu_{1}=0.499$. $E_1$ is measured independently using a TA RSA-G2 Solids Analyzer. The diameter and the impact velocity of the rubber
ball are $D=4.5$ mm and $U_{0}=0.4$ m/s, respectively. Since the surface
of the force sensor made of stainless steel is much stiffer
than the rubber ball, it barely deforms during impacts. Thus, $(1-\nu_{2}^{2})/E_{2}\ll(1-\nu_{1}^{2})/E_{1}$ and $E^* = E_1/(1-\nu_1^2)$

As an independent test, we also numerically integrate the impact force
of liquid drops over time. The resulting impulse of impacts 
\begin{equation}
J=\int_{0}^{\infty}F\mathrm{d}t
\end{equation}
quantitatively matches the momentum of impacting liquid drops before impacts (figure~\ref{fig9}\textit{b}).

\section{Propagation of isobars\label{appInertial}}
We analyze the propagation of isobars within impacting drops based on the self-similar solution of \citet{Philippi:2016bg}. The self-similar dimensionless vertical and radial coordinates are defined as $\eta\equiv z/\sqrt{\tau}$ and $\xi\equiv r/\sqrt{\tau}$. Consequently, the self-similar velocity potential and pressure fields can be written as $\widetilde{\Phi}(\eta,\xi)=\Phi(r,z,t)/\sqrt{\tau}$ and $\widetilde{P}(\eta,\xi)= \sqrt{\tau}P(r,z,t)$, respectively (cf. equations (3.1) and (3.2) in \citealt{Philippi:2016bg}). The self-similar arguments of the pressure and potential fields show that the length scale of the self-similar structure should scale as $z \sim r \sim \sqrt{\tau}$. In the dimensional form, we have $z \sim r \sim \sqrt{U_0Dt}$ as shown in $\S$\ref{subsec:Experimental-evidence}.

The self-similar pressure along the axis of symmetry ($r=0$) in the lab frame is given by (cf. eq. (3.35b) in \citealt{Philippi:2016bg})
\begin{equation}
P(r=0,z,\tau)=\frac{3\sqrt{6\tau}}{\pi\left(6\tau+4z^2\right)}. \label{eq:pressure1}
\end{equation}
Note that we use $D$, instead of $R$, as the relevant length scale to construct dimensionless variables. Thus, equation~\eqref{eq:pressure1} has different prefactors compared with equation (3.35b) in \citealt{Philippi:2016bg}.
Correspondingly, in the self-similar frame of reference, we have 
\begin{equation}
\widetilde{P}\left(0,\eta\right)=\frac{\delta}{\pi}\frac{\delta^{2}}{\delta^{2}+\eta^{2}},\label{eq:isobar}
\end{equation}
where $\delta = \sqrt{6}/2$. Far away from the impact region, Eq.~\eqref{eq:isobar} can be expanded as 
\[
\lim_{\eta\rightarrow\infty}\widetilde{P}\left(0,\eta\right)=\frac{\delta}{\pi}\left(\frac{\delta}{\eta}\right)^{2}-\frac{\delta}{\pi}\left(\frac{\delta}{\eta}\right)^{4}+\mathcal{O}\left[\left(\frac{\delta}{\eta}\right)^{6}\right].
\]
Thus, the isobar of $\widetilde{P}_0$ far from the region of impact can be obtained
through the first term of the expansion. In the self-similar frame, it is given by
\[
\frac{\delta}{\pi}\left(\frac{\delta}{\eta}\right)^{2}=\widetilde{P}_{0}
\]
This yields $\eta = \left(\delta^3/(\pi \widetilde{P}_{0})\right)^{1/2} \sim \tau^0$ in terms of the self-similar variable and $z\sim \sqrt{\tau}$ in terms of the lab-frame variable. On the other hand,
the isobar of $P_0$ in the lab frame is given by
\begin{equation}
\frac{\delta}{\pi}\left(\frac{\delta}{\eta}\right)^{2}\frac{1}{\sqrt{\tau}}=P_{0}, \label{eq:isobar1}
\end{equation}
where the additional $1/\sqrt{\tau}$ comes from the scaling of the self-similar pressure field $\widetilde{P}$. Equation~\eqref{eq:isobar1} gives the location of the isobar in the lab frame, $\eta\sim\tau^{-1/4}$ in the terms of 
self-similar variable and $z\sim\tau^{1/4}$ in terms of the lab-frame variable. Thus, in the dimensional form, we have the location of the isobar 
\[
z \sim \left(U_0D^3t\right)^{1/4} 
\]
as shown in $\S$\ref{subsec:FiniteRe}.

\section{Impact force from the asymptotic self-similar solution\label{appA}}

The self-similar shape of the drop proposed by Eggers \textit{et al.}
can be written as \citep{Lagubeau:2012ba} 
\begin{equation}
h\left(r,t\right)=h_{\max}\left(t\right)G\left(\frac{r\sqrt{h_{\max}}}{\sqrt{\Omega_{0}}}\right),
\end{equation}
where $\Omega_{0}=\pi D^{3}/6$ is the drop volume and $h_{\max}$
is the height of the drop, given by 
\begin{equation}
h_{\max}\left(t\right)=\frac{AD^{3}}{U_{0}^{2}\left(t+t_{0}\right)^{2}}.\label{eq:Lagubeaufitting}
\end{equation}
$A$ and $t_{0}$ are two fitting parameters with $A=0.492\pm0.030$
and $U_{0}t_{0}/D=0.429\pm0.033$. $G(x)$ is an unknown function
that is fixed by fitting the shape of spreading drops obtained from
either experiments \citep{Lagubeau:2012ba} or numerical simulations
\citep{Eggers:2010gd}. Notice that replacing \eqref{eq:Lagubeaufitting}
with our theoretical $h_{\max}$ from \eqref{eq:hmax} would
lead to quantitatively the same impact force shown below.

The solution of \citet{Eggers:2010gd} gives the pressure 
\begin{equation}
p(r,z,t)=\frac{3\rho\left(h\left(r,t\right)^{2}-z^{2}\right)}{\left(t+t_{0}\right)^{2}},
\end{equation}
so the force at the bottom can be obtained through integration 
\begin{eqnarray}
F(t) & = & 2\pi\int p\left(r,z=0,t\right)r\,\mathrm{d}r\\
 & = & \frac{6\pi\rho}{\left(t+t_{0}\right)^{2}}\int h^{2}\left(r,t\right)r\,\mathrm{d}r\\
 & = & \frac{6\pi\rho h_{\max}\left(t\right)\Omega_{0}}{\left(t+t_{0}\right)^{2}}\int G^{2}\left(\xi\right)\xi\,\mathrm{d}\xi.
\end{eqnarray}
Using the data shown in figure~4(\textit{a}) of \citet{Lagubeau:2012ba},
we numerically estimate the value of the integral 
\begin{equation}
I=\int G^{2}\left(\xi\right)\xi\,\mathrm{d}\xi\approx0.0911.
\end{equation}
Thus, we have the impact force 
\begin{equation}
F(t)=\frac{\rho\pi^{2}ID^{3}h_{\max}\left(t\right)}{\left(t+t_{0}\right)^{2}}=\frac{\rho\pi^{2}IAD^{6}}{U_{0}^{2}\left(t+t_{0}\right)^{4}}.
\end{equation}
We use $U_{0}=1.92$ m/s and $D=2.12$ mm, i.e., the
experimental parameters of figure~\ref{fig1}. Numerically, the corresponding impact
force is 
\begin{equation}
F(t)=\frac{10.96\textrm{ mN}\cdot\textrm{ms}^{4}}{\left(t\textrm{ ms}+0.4734\textrm{ ms}\right)^{4}},\label{eq:impactforceEggers}
\end{equation}
which is plotted as the green dotted line in figure~\ref{fig1}(\textit{b}). 

\bibliographystyle{jfm}
\bibliography{JFM_revision}

\begin{thebibliography}{49}
\expandafter\ifx\csname natexlab\endcsname\relax\def\natexlab#1{#1}\fi
\def\au#1{#1} \def\ed#1{#1} \def\yr#1{#1}\def\at#1{#1}\def\jt#1{\textit{#1}}
  \def\bt#1{#1}\def\bvol#1{\textbf{#1}} \def\vol#1{#1} \def\pg#1{#1}
  \def\publ#1{#1}\def\arxiv#1{#1}\def\org#1{#1}\def\st#1{\textit{#1}}

\bibitem[Agbaglah {\em et~al.\/}(2013)Agbaglah, Josserand \&
  Zaleski]{Agbaglah:2013jy}
{\sc \au{Agbaglah, G.}, \au{Josserand, C.} \& \au{Zaleski, S.}} \yr{2013}
  \at{{Longitudinal instability of a liquid rim}}.  \jt{Phys. Fluids}
  \bvol{25},  \pg{022103}.

\bibitem[Barenblatt(1996)]{Barenblatt:1692115}
{\sc \au{Barenblatt, G.~I.}} \yr{1996} {\em {Scaling, self-similarity, and
  intermediate asymptotics}\/}.  \publ{Cambridge, UK: Cambridge Univ. Press}.

\bibitem[Bender \& Orszag(1978)]{Orszag}
{\sc \au{Bender, C.~M.} \& \au{Orszag, S.~A.}} \yr{1978} {\em Advanced
  mathematical methods for scientists and engineers\/}.  \publ{New York, NY:
  McGraw-Hill}.

\bibitem[Bianc{\'e} {\em et~al.\/}(2006)Bianc{\'e}, Chevy, Clanet, Lagubeau \&
  Qu{\'e}r{\'e}]{Biance:2006hy}
{\sc \au{Bianc{\'e}, A.-L.}, \au{Chevy, F.}, \au{Clanet, C.}, \au{Lagubeau, G.}
  \& \au{Qu{\'e}r{\'e}, D.}} \yr{2006}  \at{{On the elasticity of an inertial
  liquid shock}}.  \jt{J. Fluid Mech.}  \bvol{554},  \pg{47--20}.

\bibitem[Brodie(1951)]{Brodie:1951cjb}
{\sc \au{Brodie, H.~J.}} \yr{1951}  \at{{The splash-cup dispersal mechanism in
  plants}}.  \jt{Can. J. Botany}  \bvol{29},  \pg{224--234}.

\bibitem[Clanet {\em et~al.\/}(2004)Clanet, B{\'e}guin, Richard \&
  Qu{\'e}r{\'e}]{Clanet:2004jg}
{\sc \au{Clanet, C.}, \au{B{\'e}guin, C.}, \au{Richard, D.} \&
  \au{Qu{\'e}r{\'e}, D.}} \yr{2004}  \at{{Maximal deformation of an impacting
  drop}}.  \jt{J. Fluid Mech.}  \bvol{517},  \pg{199--208}.

\bibitem[Deng {\em et~al.\/}(2009)Deng, Varanasi, Hsu, Bhate, Keimel, Stein \&
  Blohm]{Deng:2009t}
{\sc \au{Deng, T.}, \au{Varanasi, K.~K.}, \au{Hsu, M.}, \au{Bhate, N.},
  \au{Keimel, C.}, \au{Stein, J.} \& \au{Blohm, M.}} \yr{2009}  \at{{Nonwetting
  of impinging droplets on textured surfaces}}.  \jt{Appl. Phys. Lett.}
  \bvol{94},  \pg{133109}.

\bibitem[Dickerson {\em et~al.\/}(2012)Dickerson, Shankles, Madhavan \&
  Hu]{Dickerson:12pnas}
{\sc \au{Dickerson, A.~K.}, \au{Shankles, P.~G.}, \au{Madhavan, N.~M.} \&
  \au{Hu, D.~L.}} \yr{2012}  \at{{Mosquitoes survive raindrop collisions by
  virtue of their low mass}}.  \jt{Proc. Natl. Acad. Sci. USA}  \bvol{109},
  \pg{9822--9827}.

\bibitem[Driscoll \& Nagel(2011)]{Driscoll:2011mm}
{\sc \au{Driscoll, M.~M.} \& \au{Nagel, S.~R.}} \yr{2011}  \at{{Ultrafast
  Interference Imaging of Air in Splashing Dynamics}}.  \jt{Phys. Rev. Lett.}
  \bvol{107},  \pg{154502}.

\bibitem[Eggers {\em et~al.\/}(2010)Eggers, Fontelos, Josserand \&
  Zaleski]{Eggers:2010gd}
{\sc \au{Eggers, J.}, \au{Fontelos, M.~A}, \au{Josserand, C.} \& \au{Zaleski,
  S.}} \yr{2010}  \at{{Drop dynamics after impact on a solid wall: Theory and
  simulations}}.  \jt{Phys. Fluids}  \bvol{22},  \pg{062101}.

\bibitem[Gamero-Castano {\em et~al.\/}(2010)Gamero-Castano, Torrents, Valdevit
  \& Zheng]{Castano:2010prl}
{\sc \au{Gamero-Castano, M.}, \au{Torrents, A.}, \au{Valdevit, L.} \&
  \au{Zheng, J.-G.}} \yr{2010}  \at{{Pressure-Induced Amorphization in Silicon
  Caused by the Impact of Electrosprayed Nanodroplets}}.  \jt{Phys. Rev. Lett.}
   \bvol{105},  \pg{145701}.

\bibitem[Gart {\em et~al.\/}(2015)Gart, Mates, Megaridis \& Jung]{Gart:2015bj}
{\sc \au{Gart, S.}, \au{Mates, J.~E.}, \au{Megaridis, C.~M.} \& \au{Jung, S.}}
  \yr{2015}  \at{{Droplet Impacting a Cantilever: A Leaf-Raindrop System}}.
  \jt{Phys. Rev. Applied}  \bvol{3},  \pg{044019}.

\bibitem[Grinspan \& Gnanamoorthy(2010)]{Grinspan:2010gt}
{\sc \au{Grinspan, A.~S.} \& \au{Gnanamoorthy, R.}} \yr{2010}  \at{{Impact
  force of low velocity liquid droplets measured using piezoelectric PVDF
  film}}.  \jt{Colloid. Surface. A}  \bvol{356},  \pg{162--168}.

\bibitem[Hammitt(1980)]{Mammitt80}
{\sc \au{Hammitt, F.~G.}} \yr{1980} {\em Cavitation and multiphases flow
  phenomena\/}.  \publ{New York, NY: McGraw-Hill}.

\bibitem[Josserand \& Thoroddsen(2016)]{Josserand:2016jf}
{\sc \au{Josserand, C.} \& \au{Thoroddsen, S.~T.}} \yr{2016}  \at{{Drop Impact
  on a Solid Surface}}.  \jt{Annu. Rev. Fluid Mech.}  \bvol{48},
  \pg{365--391}.

\bibitem[Joung \& Ruie(2015)]{Joung:2015nc}
{\sc \au{Joung, Y.~S.} \& \au{Ruie, C.~R.}} \yr{2015}  \at{{Aerosol generation
  by raindrop impact on soil}}.  \jt{Nat. Commun.}  \bvol{6},  \pg{6083}.

\bibitem[Klaseboer {\em et~al.\/}(2014)Klaseboer, Manica \&
  Chan]{Klaseboer:2014ke}
{\sc \au{Klaseboer, E.}, \au{Manica, R.} \& \au{Chan, D. Y.~C.}} \yr{2014}
  \at{{Universal Behavior of the Initial Stage of Drop Impact}}.  \jt{Phys.
  Rev. Lett.}  \bvol{113},  \pg{194501}.

\bibitem[Kolinski {\em et~al.\/}(2012)Kolinski, Rubinstein, Mandre, Brenner,
  Weitz \& Mahadevan]{Kolinski:2012fx}
{\sc \au{Kolinski, J.~M.}, \au{Rubinstein, S.~M.}, \au{Mandre, S.},
  \au{Brenner, M.~P.}, \au{Weitz, D.~A.} \& \au{Mahadevan, L.}} \yr{2012}
  \at{{Skating on a Film of Air: Drops Impacting on a Surface}}.  \jt{Phys.
  Rev. Lett.}  \bvol{108},  \pg{074503}.

\bibitem[Krechetnikov \& Homsy(2009)]{Krechetnikov:2009gx}
{\sc \au{Krechetnikov, R.} \& \au{Homsy, G.~M.}} \yr{2009}  \at{{Crown-forming
  instability phenomena in the drop splash problem}}.  \jt{J. Colloid Interf.
  Sci.}  \bvol{331},  \pg{555--559}.

\bibitem[Kwon {\em et~al.\/}(2011)Kwon, Paxson, Varanasi \&
  Patankar]{Kwon:2011hm}
{\sc \au{Kwon, H.-M.}, \au{Paxson, A.~T.}, \au{Varanasi, K.~K.} \&
  \au{Patankar, N.~A.}} \yr{2011}  \at{{Rapid Deceleration-Driven Wetting
  Transition during Pendant Drop Deposition on Superhydrophobic Surfaces}}.
  \jt{Phys. Rev. Lett.}  \bvol{106},  \pg{036102}.

\bibitem[Laan {\em et~al.\/}(2014)Laan, de~Bruin, Bartolo, Josserand \&
  Bonn]{Laan:2014hj}
{\sc \au{Laan, N.}, \au{de~Bruin, K.~G.}, \au{Bartolo, D.}, \au{Josserand, C.}
  \& \au{Bonn, D.}} \yr{2014}  \at{{Maximum Diameter of Impacting Liquid
  Droplets}}.  \jt{Phys. Rev. Applied}  \bvol{2},  \pg{044018}.

\bibitem[Lagubeau {\em et~al.\/}(2012)Lagubeau, Fontelos, Josserand, Maurel,
  Pagneux \& Petitjeans]{Lagubeau:2012ba}
{\sc \au{Lagubeau, G.}, \au{Fontelos, M.~A.}, \au{Josserand, C.}, \au{Maurel,
  A.}, \au{Pagneux, V.} \& \au{Petitjeans, P.}} \yr{2012}  \at{{Spreading
  dynamics of drop impacts}}.  \jt{J. Fluid Mech.}  \bvol{713},  \pg{50--60}.

\bibitem[Landau \& Lifshitz(1986)]{LandauLifshitz:Elasticity}
{\sc \au{Landau, L.~D.} \& \au{Lifshitz, E.~M.}} \yr{1986} {\em {Theory of
  Elasticity, 3rd ed.}\/}.  \publ{Oxford, UK: Butterworth-Heinemann}.

\bibitem[Li {\em et~al.\/}(2014)Li, Zhang, Guo \& Lv]{Li:2014gq}
{\sc \au{Li, J.}, \au{Zhang, B.}, \au{Guo, P.} \& \au{Lv, Q.}} \yr{2014}
  \at{{Impact force of a low speed water droplet colliding on a solid
  surface}}.  \jt{J. Appl. Phys.}  \bvol{116},  \pg{214903}.

\bibitem[Mani {\em et~al.\/}(2010)Mani, Mandre \& Brenner]{Mani:2010m}
{\sc \au{Mani, M.}, \au{Mandre, S.} \& \au{Brenner, M.~P.}} \yr{2010}
  \at{{Events before droplet splashing on a solid surface}}.  \jt{J. Fluid
  Mech.}  \bvol{647},  \pg{163--185}.

\bibitem[Mongruel {\em et~al.\/}(2009)Mongruel, Daru, Feuillebois \&
  Tabakova]{Mongruel:2009a}
{\sc \au{Mongruel, A.}, \au{Daru, V.}, \au{Feuillebois, F.} \& \au{Tabakova,
  S.}} \yr{2009}  \at{{Early post-impact time dynamics of viscous drops onto a
  solid dry surface}}.  \jt{Phys. Fluids}  \bvol{21},  \pg{032101}.

\bibitem[Nearing {\em et~al.\/}(1986)Nearing, Bradford \&
  Holtz]{Nearing:1986da}
{\sc \au{Nearing, M.~A.}, \au{Bradford, J.~M.} \& \au{Holtz, R.~D.}} \yr{1986}
  \at{{Measurement of Force vs. Time Relations for Waterdrop Impact}}.
  \jt{Soil Sci. Soc. Am. J.}  \bvol{50},  \pg{1532--1536}.

\bibitem[Philippi {\em et~al.\/}(2016)Philippi, Lagr{\'e}e \&
  Antkowiak]{Philippi:2016bg}
{\sc \au{Philippi, J.}, \au{Lagr{\'e}e, P.-Y.} \& \au{Antkowiak, A.}} \yr{2016}
   \at{{Drop impact on a solid surface: short-time self-similarity}}.  \jt{J.
  Fluid Mech.}  \bvol{795},  \pg{96--135}.

\bibitem[Rein(1993)]{Rein:1993m}
{\sc \au{Rein, M.}} \yr{1993}  \at{{Phenomena of liquid drop impact on solid
  and liquid surfaces}}.  \jt{Fluid Dyn. Res.}  \bvol{12},  \pg{61--93}.

\bibitem[Riboux \& Gordillo(2014)]{Riboux:2014gz}
{\sc \au{Riboux, G.} \& \au{Gordillo, J.~M.}} \yr{2014}  \at{{Experiments of
  Drops Impacting a Smooth Solid Surface: A Model of the Critical Impact Speed
  for Drop Splashing}}.  \jt{Phys. Rev. Lett.}  \bvol{113},  \pg{024507}.

\bibitem[Rioboo {\em et~al.\/}(2002)Rioboo, Marengo \& Tropea]{Rioboo:2002mt}
{\sc \au{Rioboo, R.}, \au{Marengo, M.} \& \au{Tropea, C.}} \yr{2002}  \at{{Time
  evolution of liquid drop impact on solid dry surfaces}}.  \jt{Exp. Fluids}
  \bvol{33},  \pg{112--124}.

\bibitem[Roisman(2009)]{Roisman:2009cg}
{\sc \au{Roisman, I.~V.}} \yr{2009}  \at{{Inertia dominated drop collisions.
  II. An analytical solution of the Navier{\textendash}Stokes equations for a
  spreading viscous film}}.  \jt{Phys. Fluids}  \bvol{21},  \pg{052104}.

\bibitem[Roisman {\em et~al.\/}(2009)Roisman, Berberovi{\'c} \&
  Tropea]{Roisman:2009ee}
{\sc \au{Roisman, I.~V.}, \au{Berberovi{\'c}, E.} \& \au{Tropea, C.}} \yr{2009}
   \at{{Inertia dominated drop collisions. I. On the universal flow in the
  lamella}}.  \jt{Phys. Fluids}  \bvol{21},  \pg{052103}.

\bibitem[Roisman {\em et~al.\/}(2002)Roisman, Rioboo \& Tropea]{Roisman:2002iv}
{\sc \au{Roisman, I.~V.}, \au{Rioboo, R.} \& \au{Tropea, C.}} \yr{2002}
  \at{{Normal impact of a liquid drop on a dry surface: model for spreading and
  receding}}.  \jt{Proc. R. Soc. Lond. A}  \bvol{458},  \pg{1411--1430}.

\bibitem[Savic \& Boult(1955)]{Savic:1955p}
{\sc \au{Savic, P.} \& \au{Boult, G.~T.}} \yr{1955}  \at{{The fluid flow
  associated with the impact of liquid drops with solid surfaces}}.  \jt{Nat.
  Res. Council Canada}  \pg{pp. MT--26}.

\bibitem[Soto {\em et~al.\/}(2014)Soto, de~Larivi{\`e}re, Boutillon, Clanet \&
  Qu{\'e}r{\'e}]{Soto:2014fx}
{\sc \au{Soto, D.}, \au{de~Larivi{\`e}re, A.~B.}, \au{Boutillon, X.},
  \au{Clanet, C.} \& \au{Qu{\'e}r{\'e}, D.}} \yr{2014}  \at{{The force of
  impacting rain}}.  \jt{Soft Matter}  \bvol{10},  \pg{4929--4934}.

\bibitem[Tabakova {\em et~al.\/}(2012)Tabakova, Feuillebois, Mongruel, Daru \&
  Radev]{Tabakova:2012s}
{\sc \au{Tabakova, S.}, \au{Feuillebois, F.}, \au{Mongruel, A.}, \au{Daru, V.}
  \& \au{Radev, St.}} \yr{2012}  \at{{First stages of drop impact on a dry
  surface: asymptotic model}}.  \jt{Z. Angew. Math. Phys.}  \bvol{63},
  \pg{313--330}.

\bibitem[Thanh-Vinh {\em et~al.\/}(2016)Thanh-Vinh, Matsumoto \&
  Shimoyama]{Thanh-Vinh:2016n}
{\sc \au{Thanh-Vinh, N.}, \au{Matsumoto, K.} \& \au{Shimoyama, I.}} \yr{2016}
  {Pressure distribution on the contact area during the impact of a droplet on
  a textured surface}.  \bt{In {\em 2016 IEEE 29th International Conference on
  Micro Electro Mechanical Systems (MEMS)\/}},  \pg{pp. 177--180}.
  \publ{Shanghai, China: IEEE}.

\bibitem[Van~Dyke(1975)]{van1975perturbation}
{\sc \au{Van~Dyke, M.}} \yr{1975} {\em Perturbation Methods in Fluid
  Mechanics\/}.  \publ{Stanford, CA: Parabolic Press}.

\bibitem[Visser {\em et~al.\/}(2015)Visser, Frommhold, Wildeman, Mettin, Lohse
  \& Sun]{Visser:2015sm}
{\sc \au{Visser, C.~W.}, \au{Frommhold, P.~E.}, \au{Wildeman, S.}, \au{Mettin,
  R.}, \au{Lohse, D.} \& \au{Sun, C.}} \yr{2015}  \at{{Dynamics of high-speed
  micro-drop impact: numerical simulations and experiments at frame-to-frame
  times below 100 ns}}.  \jt{Soft Matter}  \bvol{11},  \pg{1708--1722}.

\bibitem[Wagner(1932)]{Wagner:1932h}
{\sc \au{Wagner, H.}} \yr{1932}  \at{{Uber stoss- und gleitvorgange and der
  oberflache von flussigkeiten}}.  \jt{Z. Angew. Math. Mech.}  \bvol{12},
  \pg{193--215}.

\bibitem[Wildeman {\em et~al.\/}(2016)Wildeman, Visser, Sun \&
  Lohse]{Wildeman:2016kj}
{\sc \au{Wildeman, S.}, \au{Visser, C.~W.}, \au{Sun, C.} \& \au{Lohse, D.}}
  \yr{2016}  \at{{On the spreading of impacting drops}}.  \jt{J. Fluid Mech.}
  \bvol{805},  \pg{636--655}.

\bibitem[Worthington(1876{\natexlab{{\em a\/}}})]{1876RSPS...25..498W}
{\sc \au{Worthington, A.~M.}} \yr{1876{\natexlab{{\em a\/}}}}  \at{{A Second
  Paper on the Forms Assumed by Drops of Liquids Falling Vertically on a
  Horizontal Plate}}.  \jt{Proc. Roy. Soc. Lond.}  \bvol{25},  \pg{498--503}.

\bibitem[Worthington(1876{\natexlab{{\em b\/}}})]{1876RSPS...25..261W}
{\sc \au{Worthington, A.~M.}} \yr{1876{\natexlab{{\em b\/}}}}  \at{{On the
  Forms Assumed by Drops of Liquids Falling Vertically on a Horizontal Plate}}.
   \jt{Proc. Roy. Soc. Lond.}  \bvol{25},  \pg{261--272}.

\bibitem[Xu {\em et~al.\/}(2005)Xu, Zhang \& Nagel]{Xu:2005dl}
{\sc \au{Xu, L.}, \au{Zhang, W.~W.} \& \au{Nagel, S.~R.}} \yr{2005}  \at{{Drop
  Splashing on a Dry Smooth Surface}}.  \jt{Phys. Rev. Lett.}  \bvol{94},
  \pg{184505}.

\bibitem[Yarin(2006)]{Yarin:2006al}
{\sc \au{Yarin, A.~L.}} \yr{2006}  \at{{Drop Impact Dynamics: Splashing,
  Spreading, Receding, Bouncing ...}}  \jt{Annu. Rev. Fluid Mech.}  \bvol{38},
  \pg{159--192}.

\bibitem[Zhang {\em et~al.\/}(2017)Zhang, Li, Guo \& Lv]{BinZhang:2017ix}
{\sc \au{Zhang, B.}, \au{Li, J.}, \au{Guo, P.} \& \au{Lv, Q.}} \yr{2017}
  \at{{Experimental studies on the effect of Reynolds and Weber numbers on the
  impact forces of low-speed droplets colliding with a solid surface}}.
  \jt{Exp. Fluids}  \bvol{58},  \pg{125}.

\bibitem[Zhao {\em et~al.\/}(2015{\natexlab{{\em a\/}}})Zhao, Zhang, Tjugito \&
  Cheng]{Zhao:2015r}
{\sc \au{Zhao, R.}, \au{Zhang, Q.}, \au{Tjugito, H.} \& \au{Cheng, X.}}
  \yr{2015{\natexlab{{\em a\/}}}}  \at{{Granular impact cratering by liquid
  drops: Understanding raindrop imprints through an analogy to asteroid
  strikes}}.  \jt{Proc. Natl. Acad. Sci. USA}  \bvol{112},  \pg{342--347}.

\bibitem[Zhao {\em et~al.\/}(2015{\natexlab{{\em b\/}}})Zhao, de~Jong \&
  van~der Meer]{Zhao:2015sm}
{\sc \au{Zhao, S.~C.}, \au{de~Jong, R.} \& \au{van~der Meer, D.}}
  \yr{2015{\natexlab{{\em b\/}}}}  \at{{Raindrop impact on sand: a dynamic
  explanation of crater morphologies}}.  \jt{Soft Matter}  \bvol{11},
  \pg{6562--6568}.

\end{thebibliography}

\end{document}